\documentclass[11pt,a4paper]{article}
\usepackage{jcappub}
\usepackage{amsmath, amssymb, amsthm, epsfig, fancyhdr,epsfig,multirow}
\usepackage[top=1in, bottom=1in, left=1.3in, right=1.3in]{geometry}
\usepackage[utf8]{inputenc}
\usepackage{amsmath}
\usepackage{amsfonts}
\usepackage{amssymb}
\usepackage{xcolor}
\usepackage{comment}
\usepackage[multiple]{footmisc}
\usepackage{subfig}
\usepackage[normalem]{ulem}
\usepackage{tabularx}
\usepackage{comment}
\usepackage{float}
\usepackage{graphicx}
\usepackage{appendix}
\usepackage[most]{tcolorbox}
\usepackage{physics}
\usepackage{mathtools}
\title{Singularities in Cosmological Loop Correlators II : Non Local Interactions and Flat Space limits}
\usepackage{xcolor}
\usepackage{tikz}
\author[1]{Elaf Ansari,}
\author[1]{Supritha Bhowmick,}
\author[1]{Diptimoy Ghosh}
\affiliation[1]{Indian Institute of Science Education and Research Pune, Pune 411008, India}

\emailAdd{ansari.mohdelaf@students.iiserpune.ac.in}
\emailAdd{supritha.bhowmick@students.iiserpune.ac.in} 
\emailAdd{diptimoy.ghosh@gmail.com} 

\abstract{Non-local interactions naturally arise in the ADM formalism after solving the constraint equations and substituting their solutions back into the action. However, the effects of these non-local operators on loop corrections to cosmological correlators remain largely unexplored. Extending the analysis of \cite{Bhowmick:2025mxh}, in this work we compute the $1$-loop bispectrum during slow-roll inflation, including the inverse-Laplacian operators present at cubic and quartic order in the ADM action (in the spatially flat gauge). These non-local vertices introduce non-trivial angular dependencies that significantly complicate the evaluation of loop integrals. We find that the diagrammatic rules of \cite{Bhowmick:2025mxh} for identifying poles and branch cuts in correlators without explicit integration remain valid even in the presence of such non-local interactions. Finally, we derive the flat-space limit of the $1$-loop inflationary correlator and verify it with explicit examples, thereby establishing a direct correspondence between its leading total-energy singularities ($\omega_T\rightarrow0$) and the flat-space scattering amplitude.}

\begin{document}
\maketitle
\flushbottom

\section{Introduction}
Inflation remains the leading paradigm for explaining the early universe’s quantum origin of structure. It accounts for the observed flatness, homogeneity, and isotropy of the universe, as well as seeding the inhomogeneities observed in the cosmic microwave background. Considerable progress has been made in understanding the correlations of inflationary perturbations at tree level (for a representative set of references, see \cite{Maldacena:2002vr,Arkani-Hamed:2018kmz,Baumann:2019oyu,Baumann:2020dch,Pajer:2020wnj,Pajer:2020wxk,Jazayeri:2021fvk,DuasoPueyo:2020ywa,Ghosh:2022cny,Ghosh:2023agt,Sleight:2019mgd,Sleight:2019hfp,Arkani-Hamed:2015bza,Arkani-Hamed:2017fdk,Baumgart:2019clc}), which map directly onto cosmological observables. However, to rigorously test the robustness of inflationary theories and their predictions, it is essential to incorporate quantum loop corrections into these correlators.

Loop effects in cosmology are generally small (being higher order in perturbation theory), but they play a crucial role in several contexts. They serve as a nontrivial check on the reliability of the perturbative expansion and can alter predictions for observables such as the level of non-Gaussianity \cite{Riotto:2008mv,Assassi:2012et,Gorbenko:2019rza,Green:2020txs,Cohen:2021fzf,Wang:2021qez,Lee:2023jby,Maru:2021ezc}. Loop corrections have also been studied in the context of secular infrared growth of cosmological correlations \cite{Ford:1984hs,Antoniadis:1985pj,Starobinsky:1994bd,Dolgov:2005se,Marolf:2010zp,Burgess:2010dd, Tsamis:2005hd, Prokopec:2007ak,Woodard:2025cez,Woodard:2025smz,Foraci:2024cwi,Foraci:2024vng,Miao:2024shs,Cespedes:2023aal,Miao:2025gzm} and even in scenarios like the formation of primordial black holes \cite{Sasaki:2018dmp,Kristiano:2022maq,Riotto:2023hoz,Kristiano:2023scm,Riotto:2023gpm,Choudhury:2023jlt,Franciolini:2023agm}. More broadly, analyzing loop corrections can shed light on how fundamental principles — unitarity \footnote{The consequence of unitarity in tree level wavefunction coefficients has been studied in \cite{Goodhew:2020hob,Jazayeri:2021fvk,Melville:2021lst,Goodhew:2021oqg,Ghosh:2024aqd,Ghosh:2025pxn}}, locality, and causality — manifest in the analytic structure of inflationary correlators, much as they do in the singularity structure of flat-space scattering amplitudes \cite{Eden:1966dnq, Martin:1969ina,Elvang:2015rqa,Nussenzveig:1972tcd,Sashanotes, Mizera:2023tfe,Adams_2006,Azatov_2022,Ghosh:2022qqq,Desai:2025alt,de_Rham_2017,de_Rham_2018,Bellazzini_2021,Caron_Huot_2021,Tolley_2021,Sinha_2021,Henriksson_2022,Riembau:2022yse,Arkani_Hamed_2021,Chiang:2022ltp}. Unlike flat-space amplitudes, however, cosmological loop correlators remain relatively uncharted territory. Prior to recent work, little was known about their analytic structure. Some progress has come from studying singularities of the cosmological wavefunction (or wavefunction coefficients) at $1$-loop order \cite{Benincasa:2024ptf,Chowdhury:2023khl,Beneke:2023wmt,Goodhew:2024eup}, but the singularity structure of actual observables (correlation functions) can differ markedly \cite{Chowdhury:2023arc,Chowdhury:2025ohm,Westerdijk:2025ywh,Cacciatori:2024zrv} due to different physical conditions (such as the choice of vacuum state or boundary conditions in the inflationary past).

Recently, a companion analysis (Ref.~\cite{Bhowmick:2025mxh}) provided a set of diagrammatic rules to efficiently extract the singularity structure of 1-loop inflationary correlators. These rules make it possible to identify poles and branch cuts in a correlator by examining sub-diagrams and energy flows, without evaluating the full loop integrals. However, those results — and indeed all known results on loop correlator analytic structures so far — were derived for local interactions, i.e. polynomial or derivative couplings of the form $(\partial\phi)^n$ or $\phi^n$. Extending beyond simple local interactions is challenging. Generalizing the diagrammatic rules beyond $1$-loop diagrams with two interaction vertices (the “bubble” topologies) leads to significantly more complex analytic structures. In fact, the only analytic result for the next-nontrivial one-loop topology (the triangle diagram) has been obtained in the context of the cosmological wavefunction \cite{Benincasa:2024ptf}. No intuition or results existed for how more general, non-local interaction terms affect the singularity structure of loop corrections. This is an important gap, because realistic inflationary effective actions naturally contain non-local interactions. For example, when using the ADM formalism to enforce the Hamiltonian and momentum constraints, one finds inverse-Laplacian operators appearing in the cubic \cite{Maldacena:2002vr} and quartic \cite{seery2007inflationary,bonifacio2023graviton} Lagrangians. An illustrative term is a cubic interaction of the form $S_3\sim\int\dot{\zeta}^2 \partial^{-2} \dot{\zeta}$ (involving an inverse spatial Laplacian). Such non-local vertices introduce non-trivial angular dependencies in loop integrals and can significantly complicate their evaluation. It is therefore both technically interesting and phenomenologically important to understand how these non-local interactions influence the analytic structure of cosmological loop correlators.

In this work, we address this question by extending the analysis of Ref.~\cite{Bhowmick:2025mxh} to include non-local inflationary interactions and studying their impact on the $1$-loop bispectrum. We focus on a representative set of non-local terms from the cubic and quartic action in the spatially flat gauge (arising from solving the ADM constraints) and compute the corresponding 1-loop corrections to the scalar bispectrum. Our calculation is performed using two complementary regularization schemes – dimensional regularization and a physical momentum cutoff – to cross-check results and handle the various divergences. The inverse-Laplacian factors in the interaction vertices lead to highly non-trivial momentum dependence (for instance, momentum dot products in the numerator of interaction factors), which renders some loop integrals analytically intractable in general $d$ dimensions. In those cases we evaluate the problematic contributions using the cutoff regulator. To our knowledge, this work provides the first analysis of the analytic structure of loop-level cosmological correlators including interactions with inverse-Laplacian (non-local) operators. Rather than attempt an exhaustive computation of all possible interaction terms, our goal is to isolate and explore the singularity signatures produced by a few salient non-local interactions from the ADM action. This approach allows us to clearly identify which features in the $1$-loop bispectrum arise from the non-local nature of the vertices.

Another key aspect of our study is the investigation of the flat-space limit for the $1$-loop bispectrum. It is well-known that for tree-level cosmological correlators, taking the total energy $\omega_T$ of the external fields to zero relates the leading singularity of the correlator to a flat-space scattering amplitude describing the same process \cite{Maldacena:2011nz,Raju:2012zr,Arkani-Hamed:2017fdk,Arkani-Hamed:2018kmz,Benincasa:2018ssx,Baumann:2020dch}. In particular, the flat-space scattering amplitude can be recovered from the highest-order pole (as $\omega_T \to 0$) of the tree-level inflationary correlator \cite{Goodhew:2020hob}, and this correspondence extends to the integrands of loop diagrams. However, the flat-space limit of the full loop-corrected correlator (after performing the loop integration) has not been derived before. We fill this gap by explicitly deriving the flat-space limit of the complete $1$-loop bispectrum (with non-local interactions included) and verifying it in concrete examples. We find that, as expected, the leading singular part of the 1-loop correlator reproduces the corresponding flat-space scattering amplitude. At the same time, the expanding de Sitter background introduces additional features in the loop correlator — notably total-energy branch cuts — which have no direct analogue in flat space. We show that in the renormalized 1-loop bispectrum, all such branch-cut contributions organize into dilatation-invariant logarithms of momentum ratios, similar to what was observed in Ref.~\cite{Bhowmick:2025mxh} for loops with local interactions. Interestingly, we also find that these extra de Sitter features manifest differently depending on the regularization scheme. In both dimensional and cutoff regularization the same on-shell scattering amplitude is obtained in the $\omega_T \to 0$ limit, but the two schemes yield slightly different residual logarithmic terms coming from the total-energy branch cuts. We discuss these scheme-dependent differences and their interpretation in Section \ref{Sec:Polology}.

We summarize the main results of this paper as follows:
\begin{itemize}
    \item  \textbf{Singularities with non-local interactions:} The diagrammatic rules from Ref.~\cite{Bhowmick:2025mxh} for predicting singularity structure at one-loop remain valid even in the presence of non-local (inverse-Laplacian) interaction vertices. We explicitly verify that the pole and branch cut positions inferred by these rules match the results of our direct $1$-loop calculations with non-local terms.
    \item \textbf{Branch cuts and logarithmic terms:} Loop corrections involving non-local interactions exhibit branch cuts in the total-energy variable. In the renormalized correlator, all such branch cuts repackage into logarithmic terms that respect dilatation (scale) invariance. This behavior is analogous to what was found in the simpler case of local interactions \cite{Bhowmick:2025mxh}, indicating a degree of universality in how de Sitter symmetry constrains loop correlators.
    \item \textbf{Flat-space limit at $1$-loop:} In the limit $\omega_T \to 0$, the leading singular part of the $1$-loop bispectrum with non-local interactions reduces to the corresponding flat-space scattering amplitude, in both dimensional and cutoff regularization schemes. However, due to the time-dependent de Sitter background, the $1$-loop correlator also contains features with no flat-space analogue (arising from total-energy branch cuts). These extra contributions manifest differently in the two regularization schemes, resulting in slight scheme-dependent differences in the exact flat-space limit for the branch-cut pieces. Importantly, the physical scattering amplitude itself is consistent between schemes – the scheme differences only affect the additional dS-specific terms (See Sec.~\ref{Sec:Polology}).
\end{itemize}

Finally, the paper is organized as follows. In Section \ref{Sec:Computational_Scheme} we outline our computational approach for evaluating $1$-loop bispectra with both cutoff and dimensional regularization. Section \ref{Sec:Explicit_Comp_Bispectrum} presents the explicit 1-loop bispectrum results obtained from various non-local interactions, and Section \ref{Sec:Complicated_Interactions} discusses additional contributions and cross-checks (including interactions omitted from the main text). In Section \ref{Sec:Polology}, we examine the flat-space limit of the $1$-loop correlator in detail and derive the relationship between the singularities of the de Sitter correlators and the flat-space scattering amplitude. We conclude with a summary of results and future outlook in Section \ref{Sec:Conclusion}.

\section{Notations and Conventions}
\label{Sec:Notations}
The metric in the spatially flat gauge is given by,
\begin{align}
ds^2 = -N^2 dt^2 + e^{2\rho(t)} (dx^i + N^i dt)(dx^j + N^j dt) \,,
\end{align}
where $\dot{\rho}(t)$ is the Hubble parameter ($H$), considered to be time independent here to leading order in the slow-roll parameters. The constraints : the lapse function $N$ and the shift vector $N_i$ are evaluated to second order to get the action till fourth order in perturbation \footnote{higher order terms in the constraints multiply terms that vanish since they satisfy equation of motion; see Appendix  of \cite{Pajer_2017}}. Since the constraint equations are of second order in spatial derivatives, the constraint solutions will generally contain inverse laplacian, giving non-local terms in cubic and quartic actions. The background and fluctuations of the scalar field is denoted by,
\begin{align}
    \phi(x,t)=\bar{\phi}(t)+\varphi(x,t) \,.
\end{align}

Focusing on the scalar sector, the quadratic and cubic action at leading order in slow roll in the spatially flat guage is given by \cite{Maldacena:2002vr},
\begin{align}   
&S_2 = \frac{1}{2} \int dt \, d^3x  
\left[ e^{3\rho} \dot{\varphi}^2 - e^{\rho} (\partial \varphi)^2 \right] \,, \nonumber \\
&S_3 = \int dt d^3x e^{3\rho} \left( 
- \frac{\dot{\bar\phi}}{4\dot{\rho}} \varphi \dot{\varphi}^2 
- e^{-2\rho} \frac{\dot{\bar\phi}}{4\dot{\rho}} \varphi (\partial \varphi)^2 
-\dot{\varphi} \partial_i \beta \partial_i \varphi
\right) \,, \label{Eqn:cubic_action}
\end{align}

where
\begin{align}
    \partial^2 \beta =\left[ \frac{\dot{\bar{\phi}}^2}{2 \dot{\rho}^2} \frac{d}{dt} \left( - \frac{\dot{\rho}}{\dot{\bar\phi}} \varphi \right) \right] \,.  \label{Eqn:cubic_action_constraint}
\end{align}
The quartic action (leading order in slow-roll), computed in \cite{seery2007inflationary,bonifacio2023graviton} is given by,
\begin{align}
    S_4 = &\frac{1}{4} \int dt \, d^3x \, e^{3\rho} 
\bigg[ 
\partial^{-2}\partial_i (\dot\varphi \partial_i \varphi) 
\partial^{-2} \partial_j (\dot\varphi \partial_j \varphi) 
- \frac{1}{\dot{\rho}} \partial^{-2} \partial_j (\dot\varphi \partial_j \varphi) 
\left( \dot{\varphi}^2 + e^{-2\rho} \partial_i \varphi \partial_i \varphi \right) \nonumber \\
&+ 4 \dot\varphi \partial_i \varphi \partial^{-2} (\dot\varphi \partial_i \varphi) 
\bigg].  \label{Eqn:quartic_action}
\end{align}

Note that since the interactions contain temporal derivative of fields, one should be careful when evaluating the interaction Hamiltonian, since legendre transforming the lagrangian will now generate higher order operators. However, since we work in leading order in slow-roll, the quartic terms generated from the cubic interaction can be neglected (see Appendix \ref{App_Time_dep_Hamilt}).

In our computations, we denote the external momenta by $\{\vec{k}_i\}$ and internal legs by $\{\vec{p}_i\}$, with $\vec{p}_1$ \& $\vec{p}_2$ reserved for loop. The magnitude of the momentum exchanged between the loop sites is denoted by $s=|\vec{p}_1+\vec{p}_2|$.

For \textit{on-shell} propagators, the energy of the propagator is defined as the modulus of the momentum, i.e. $k_i=|\vec{k}_i|$, while for \textit{off-shell} propagators, the energy is treated as independent variable, denoted by $\{\omega_i\}$. This distinction is useful to define the \textit{off-shell correlator}, obtained as an analytic extension of the physical correlator, where the energy of the internal legs remains \textit{on-shell}, i.e. $p_i=\vec{p}_i$, but the energy of the external legs $\omega_i$ is treated as independent from the momentum $\vec{k}_i$. Physical answers are obtained when we put the external legs \textit{on-shell}, i.e. take $\omega_i=|\vec{k}_i|$.
The purpose of this analytic continuation is to make transparent the singularity structure of correlators \cite{Salcedo:2022aal,AguiSalcedo:2023nds,Lee:2023kno}, which allowed a systematic analysis of singularity structure for 1-loop diagrams, leading to diagrammatic rules for derivative and polynomial interactions \cite{Bhowmick:2025mxh}. In this work, we will compute off-shell correlators to inspect whether the diagrammatic rules apply to non-local interactions as well. Finally, we will denote sum of off-shell energies by writing the indices one after the other in the subscript, such as $\omega_1+\omega_2+\omega_5=\omega_{125}$ and so on.

\section{Details of Computational Scheme}
\label{Sec:Computational_Scheme}
Unlike collider experiments, where initial and final states are well-defined, cosmological observations involve measurements at a single final time slice. To compute expectation values in this context, we employ the \textit{in-in} formalism. The expectation value of an operator \(\mathcal{O}(\tau_0)\) at time $\tau_0$ is given by

\begin{equation}
    \langle \mathcal{O}(\tau_0) \rangle = \left\langle 0 \left| \bar{T} e^{i \int_{-\infty(1+i \epsilon)}^{\tau_0} H_I d\tau} \mathcal{O}(\tau_0) T e^{-i \int_{-\infty(1-i\epsilon)}^{\tau_0} H_I d\tau} \right| 0 \right\rangle \,.
\end{equation}

Here, \(T\) and \(\bar{T}\) denote time and anti-time ordering, respectively, \(H_I\) is the interaction Hamiltonian and $\ket{0}$ is a free vacuum state. We will compute the correlators at future boundary of dS, i.e. $\tau_0\rightarrow0$. The structure of this expression gives rise to diagrams, often called \textit{Feynman-Witten diagrams} \cite{chen2017schwinger}, with vertices drawn below (called ``right" or ``+") or above (called ``left" or ``-") $\tau=0$ line, depending on whether they come from the time ordered or the anti-time ordered part of the correlator respectively (following notation of \cite{Weinberg:2005vy, Chen:2017ryl}). Not all of these diagram need to be computed since half of these combinations are obtained from complex conjugation of the other half (provided the Hamiltonian is hermitian). The propagators joining two bulk points are given by,
\begin{align}
    G_{++}(k;\tau_1,\tau_2)&=f_k^\ast(\tau_1)f_k(\tau_2)\Theta(\tau_1-\tau_2)+f_k(\tau_1)f_k^\ast(\tau_2)\Theta(\tau_2-\tau_1),\\
    G_{+-}(k;\tau_1,\tau_2)&=f_k(\tau_1)f^\ast_k(\tau_2),\\
    G_{--}(k;\tau_1,\tau_2)&=G_{++}^\ast(k;\tau_1,\tau_2),\\
     G_{-+}(k;\tau_1,\tau_2)&=G_{+-}^\ast(k;\tau_1,\tau_2), \label{Eqn: Propagators}
\end{align}
where $+$ and $-$ in the subscripts denote whether the interaction vertices $\tau_1,\tau_2$ lie below or above $\tau=0$ line respectively. Depending on the interactions of the theory, the modes in the propagators will be acted upon by operators (spatial/time derivatives or inverse laplacians).

We will compute 1-loop diagrams with two regularization schemes (dimensional regularization and cutoff regularization) wherever possible. For certain contractions, the integral in dim reg becomes intractable due to complicated vertex factors in $d$-dimensions coming from the non local interactions; for such cases we give the answer from cutoff regularization only. 

\paragraph{Dimensional Regularization : }
Mode functions in de-Sitter are dimension-dependent unlike in flat space. This requires considering full $d=3+\delta$ dimensional modes when computing loop integrals in dim reg, with $\mathcal{O}(\delta)$ and higher order corrections from the modes contributing crucially to the structure of the correlator. This is well known from previous 1-loop computations \cite{Weinberg:2005vy,Adshead:2008gk,Chaicherdsakul:2006ui,Seery:2007we,Gao:2009fx}, where the unphysical logarithms were tamed by properly regulating the mode functions (\cite{Senatore:2009cf}) and renormalising the correlator (\cite{Bhowmick:2024kld}).

In de Sitter, the mode function in $d$-dimensions for a scalar field with mass $m$ is given by:
\begin{equation}
    f_k(\tau)=H^{-\frac{1}{2}}(-H\tau)^{d/2}H^{(2)}_{i\nu}(-k\tau),
\end{equation}
where $\nu=\sqrt{\frac{m^2}{H^2}-\frac{d^2}{4}}$. However in arbitrary $d$-dimension, integral involving these Hankel functions are complicated. Therefore, we analytically continue the mass as \cite{Melville:2021lst,Lee:2023jby,Jain:2025maa},
\begin{align}
    \frac{m^2}{ H^2} \to \left(\frac{d^2}{4} - \frac{9}{4} \right) \,,
\end{align}
which gives back simple massless modes with $\nu=3/2$. The time integral measure also receives correction in dimensional regularization. Thus the $d$-dimensional modes and measure simplify as,
\begin{align}
    &f_k(\tau) = (-H\tau)^{\delta/2}  \frac{H}{\sqrt{2k^3}} (1 - ik\tau) e^{ik\tau} = (-H\tau)^{\delta/2} f_k^{3d}(\tau) \,,  \nonumber \\
    &a^\delta (\tau)=(-H\tau)^{-\delta}\,, \label{Eqn:modes_d_dim}
\end{align}
where $f_k^{3d}(\tau)$ are $3$-dimensional dS modes. In practice, the above is expanded about $\delta=0$ till $\mathcal{O}(\delta)$ (since 1-loop corrections are at most $\frac{1}{\delta}$ divergent). This gives,
\begin{align}
     &f_k(\tau)=\bigg(1+\frac{\delta}{2}\log(-H\tau)\bigg)f_k^{3d}(\tau)+\mathcal{O}(\delta^2)  \nonumber \\ 
     &a^\delta (\tau)=1-\delta \log(-H \tau) +\mathcal{O}(\delta^2)\,.
     \label{Eqn:modes_d_dim_expand}
\end{align}

We will now sketch the calculational scheme in dim reg. In our calculations, we treat the external mode functions at \( \tau = 0 \) and the momentum-conserving delta function \( \delta^d \left( \sum_i \vec{k}_i \right) \) as three-dimensional, since the counterterms precisely cancel the $\mathcal{O}(\delta)$ corrections arising from these terms \cite{Senatore:2009cf}.

The general expression for one loop correction to the \textit{off-shell} bispectrum (Fig.~\ref{Fig:Bispectrum}) \( D(\vec{k}_i,\omega_i) \) has the general form,
\begin{align}
D(\vec{k}_i,\omega_i) = &\delta^3 \left(\sum_i \vec{k}_i \right) 
\mu^{- 3\delta/2}
\int d^d \vec{p}_1 \, d^d \vec{p}_2 \, Q(p_1, p_2, \{ \omega_i \},\{\vec{k_j}\cdot \vec{p}_1 \}, \{ \tau_l \}) \delta^d (\vec{p}_1 + \vec{p}_2 - \vec{s}) \nonumber \\
& 
\times \prod_l \left[ (-H \tau_l)^{n_l \delta} d\tau_l \right]\,,
\end{align}
where the factor of the renormalisation scale $\mu$ has been introduced to account for the $\delta$ dependence in the mass dimensions of the couplings in $d$-dimensions. $\vec{s}(=\vec{k}_3=-\vec{k}_1-\vec{k}_2)$ is the momentum exchanged between the loop sites. $\tau_i$ are bulk vertices which are integrated over, $p_1,p_2$ are loop momenta, and $Q$ is a rational function of $\tau_i$'s, external and loop momenta. Contributions from vertex factors involving dot products (arising from spatial derivatives in the interaction) or inverse square norms (from non local operators) of momenta have been absorbed in $Q$. Also note that $Q$ consists only of the $3$-d contribution from modes and measure in Eqn.~\ref{Eqn:modes_d_dim}. The $\mathcal{O}(\delta)$ contribution from modes and measure has been extracted in the second line in the above equation, with $n_i=\frac{1}{2},-1$ when considering contribution from modes and measure respectively. Note $\sum_l n_l$ equals the power of $\mu$ for any diagrams, this is by construction due to how powers of $\mu$ are introduced in the correlator, and hence we get terms like $\log \bigg(\frac{H}{\mu} \bigg)$ at 1-loop. 

Expanding the last term using Eqn.~\eqref{Eqn:modes_d_dim_expand} we get,
\begin{align}
D(\vec{k}_i,\omega_i) = &\delta^3 \left(\sum_i \vec{k}_i \right) 
\mu^{- 3\delta/2}
\int d^d \vec{p}_1 \, d^d \vec{p}_2 \,Q(p_1, p_2, \{ \omega_i \},\{\vec{k_j}\cdot \vec{p}_1 \}, \{ \tau_l \})\delta^d (\vec{p}_1 + \vec{p}_2 - \vec{s}) \nonumber \\
& 
\times \prod_i  \left( 1 + \sum_i n_i \delta \log(-H\tau_i)\right) d\tau_i \,,
\end{align}
where the first term in the second line can be identified as $D_1$- the $d$-dimensional loop integrals with $3$-dimensional modes, given by
\begin{align}
    D_1(\vec{k}_i,\omega_i) = &\delta^3 \left(\sum_i \vec{k}_i \right) 
\mu^{- 3\delta/2}
\int d^d \vec{p}_1 \, d^d \vec{p}_2 ~Q(p_1, p_2, \{ \omega_i \},\{\vec{k_j}\cdot \vec{p}_1 \}, \{ \tau_l \})\nonumber \\
&\delta^d (\vec{p}_1 + \vec{p}_2 - \vec{s})\prod_i  d\tau_i \,, \label{Eqn:D_1}
\end{align}
and the second term, which we call $D_2$, is the contribution to the loop diagram coming from $\mathcal{O}(\delta)$ corrections from modes and measure. Clearly the only difference in the computations of $D_2$ and $D_1$ is that the former has additional factors of $\log (-H \tau)$ in time integrals. For local interactions, this factor can be simply pulled out of the time integrals by substituting $\log(-\tau)\rightarrow \log(1/\omega_T)$  \cite{Senatore:2009cf,Bhowmick:2024kld} (upto some non-logarithmic finite functions of $\omega_i$). In Appendix.~\ref{App_D_2}, we show that the same relation,
\begin{align}
    D_2(\vec{k}_i,\omega_i)=\delta~ D_1\sum_{i}n_i\log(H/\omega_T) + \text{~finite non-log}\,, \label{Eqn:D_2}
\end{align}
also holds for the computations in this work involving inverse laplacian interactions.

To evaluate $D_1$, it is convenient to compute the time integrals first. With increasing number of interaction vertices, one needs to compute more and more time orderings, hence this calculation is long but straightforward. Next, the loop integrals must be evaluated. In the presence of non-local operators, this can be tricky. These loop integrals are evaluated using an identity which convert $\int d^d\vec{p}_1 d^d \vec{p}_2 \rightarrow \int dp_1 dp_2$, for e.g. see Appendix A of \cite{Bhowmick:2024kld} for loop diagrams involving local interactions. To briefly sketch the idea- the $d$-dimensional integral over $\vec{p}_2$ is carried out using the momentum conserving delta function $\delta^d(\vec{p}_1+\vec{p}_2-\vec{s})$. Next, the integral over $p_1$ is split into norm and angular variables as,
\begin{align}
    \int d^d \vec{p}_1 \rightarrow \int p_1^{d-1} \sin^{d-2}{\theta_1} \sin^{d-3}{\theta_2}...\sin{\theta_{d-2}} ~ ~dp_1 d\theta_1 d\theta_2 d\theta_{d-2} d\phi   \,, \label{Eqn:phase_space_int_split}
\end{align}
where $\phi$ is the azimuthal angle and $\theta_i$'s denote polar angles. With $\hat{s}$ chosen along the axis with respect to which $\theta_1$ is measured, we have,
\begin{align}
    |\vec{s}-\vec{p}_1| =\sqrt{s^2 +p_1^2-2s p_1 \cos{\theta_1}}=p_2 \,.
\end{align}
Now the rest of the derivation proceeds by performing a variable change from $\theta_1$ to $p_2$ and integrating the chain of $\sin$ factors over $d\theta_2 ... d \theta_{d-2}d\phi$ as $\int d\Omega_{d-1}$. This variable change and integral is trivial if the interactions involve only spatial derivatives, where the vertex contribution involves factors such as,
\begin{align}
    \vec{k}_1 \cdot \vec{p}_1 = k_1 p_1 \left[ \cos \theta_1 \cos \theta_k + \sin \theta_1 \sin \theta_k \cos \theta_2 \right] \,,  \label{eqn:spatial_Der_mom_identity} 
\end{align}
where $\vec{k}_1$ is chosen to lie in the plane of $\hat{s}$ and the axis with respect to which $\theta_2$ is measured, $\theta_k$ is angle between $\vec{k}_1$ and $\vec{s}$, given by $\cos \theta_k = \frac{k_2^2-k_1^2-s^2}{2k_1s}$. Importantly, we only need to keep the first term since the second term in the integrand is odd over $\theta_2 \in (0,\pi)$ due to $\cos \theta_2$.

However, on top of spatial derivatives, we now have inverse laplacian operators acting on the fields. This can give rise to vertex factors such as,
\begin{align}
     \frac{\vec{p}_1\cdot\big(\vec{k}_2 + \vec{p}_1 \big)}{|\vec{k}_2+\vec{p}_1|^2} \xrightarrow{\text{off-shell}}\frac{ \omega_2 p_1 \big( \cos \theta_1 \cos \theta_k + \sin \theta_1 \sin \theta_k \cos \theta_2 \big) + p_1^2}{\omega_2^2 + p_1^2 +2 \omega_2 p_1 \big( \cos \theta_1 \cos \theta_k + \sin \theta_1 \sin \theta_k \cos \theta_2 \big)} \,,  \label{Eqn:complicated_vertex}
\end{align}
where we considered the vertex factor in Diagram 31 (see Section.~\ref{Sec:Explicit_Comp_Bispectrum}), and $k_i$'s are replaced with $\omega_i$ in the offshell limit. $\theta_k$ is angle between $\vec{k}_2$ and $\vec{s}$. Now the integral over angular variables is no longer tractable due to terms such as $\sin^{d-3} \theta_2$. Fortunately, this calculation can still be carried out in $d=3$, thus such contractions are to be handled in cutoff regularization only. In Section.~\ref{Sec:Explicit_Comp_Bispectrum}, we notice that the angular dependence can be trivialized in the contractions where both the fields, on which inverse laplacian operator acts, contract to form internal legs, or both contract to form external legs. For e.g. in Diagram 12, the vertex contribution is given by,
\begin{align}
    \frac{\vec{p}_1 \cdot \big(\vec{p}_1+\vec{p}_2 \big)}{|\vec{p}_1+\vec{p}_2|^2} =\frac{\vec{p}_1 \cdot \vec{s}}{s^2}=\frac{s^2+p_1^2-p_2^2}{s^2} \,, \label{eqn:inv_laplac_Der_mom_identity}
\end{align}which introduces no dependence on $\theta_i$ ($i=1,...,d-2$) or $\phi$. For such contractions, we compute the diagram in both- dim reg as well as cutoff regularization. In such cases, using the loop integral identity we find Eqn.~\eqref{Eqn:D_1} gives,
\begin{align}
          D_1(\vec{k}_i,\omega_i)=\delta^3\big(\sum_i \vec{k}_i \big)~\mu^{-\frac{3 \delta}{2}}  \frac{S_{d-2}}{2}\int_{s}^{\infty}dp_{+} \int_{-s}^{s} dp_{-} \frac{p_1^{d-2}p_2}{s} P(p_{\pm},\{\omega_i\},s) \sin^{\delta}\theta_1 \,.
    \end{align}
where $S_{d-2}$ is surface area of a $(d-2)$ - unit sphere, $P$ is a rational function of $p_{\pm},s$ and external energies, and we have defined $p_{\pm}=p_1\pm p_2$. Note that although the factor of $\sin^{\delta}\theta_1$ does not contribute to the $\mathcal{O}(\frac{1}{\delta})$ divergent terms \cite{Melville:2021lst} in the correlator, it does give finite contributions and hence is important in determining the structure of the correlator.

\paragraph{Cutoff Regularization:}
As previously explained, in the presence of non-local interactions, the vertex factors for some contractions become too complicated to be handled in $d$-dimensions. In such a scenario, we compute the integral in cutoff regularisation only. In $d=3$, all $\theta_i$'s drop out of Eqn.~\eqref{Eqn:phase_space_int_split} for $i>1$, and the integrand simplifies enormously, making the calculation tractable even in the presence of complicated vertex factors such as Eqn.~\eqref{Eqn:complicated_vertex}. 

There is another reason for performing computations in cutoff, even for cases when calculation in dim reg is possible. Note that the computations performed in dimensional regularization involved commutation of a summation operator (coming from the Taylor series expansion of the modes about $\delta=0$) across the loop integrals. This commutation is subtle - for the integral to converge, $\delta$ must be analytically continued to some point in the complex plane, and arguing the convergence of the sum of integrals is not straightforward at this value of $\delta$. Thus, a cross check from cutoff regularization demonstrates that answers from dimensional regularization still reliably compute the logarithmic terms, which are the relevant branch cuts we are looking to explore in these correlators.

In cutoff regularization, we regulate the momentum integrals by placing a cutoff at $\Lambda a(\tau_f)$, where $\Lambda$ is the physical cutoff and $\tau_f$ is the time being integrated first, e.g. for the $\Theta(\tau_1 -\tau_2)$ part of the integrand $\tau_f =\tau_2$. We integrate the loop integrals first, since the cutoff is now time dependent. The loop momentum identity now becomes : \begin{align}
    & \int d^3 \vec{p}_1 d^3 \vec{p}_2 Q(p_1,p_2,\{\omega_i\},\{\tau_l\}) \delta^3 (\vec{p}_1 + \vec{p}_2 +\vec{k}_3) \mathcal{Q}(p_1,\{ \omega_i \},\{\vec{k}_j \cdot \vec{p}_1\}) \nonumber \\
    &= \int_0^{\Lambda a(\tau_f)} dp_1 ~p_1^2 Q(p_1,p_2,\{\omega_i\},\{\tau_l\}) \int d\theta d\phi ~\sin \theta ~\mathcal{Q}(p_1,\{\omega_i\},\theta,\phi)  \,, \label{Eqn:cutoff_vertex_complicated}
\end{align}
where we separate the vertex factor from $Q$ and write it as $\mathcal{Q}$. For inverse laplacian operators acting on fields $\big(\varphi_1,\varphi_2\big)$, the $\phi$ dependence of $\mathcal{Q}$ remains trivial as long as both the modes of $\varphi_1,\varphi_2$ contract in external or internal legs (see Eqn.~\eqref{eqn:inv_laplac_Der_mom_identity}). In this case changing variables from $\theta$ to $p_2$ is straightforward. However the structure of $\mathcal{Q}$ gets significantly more complicated when one of $\varphi_1$ or $\varphi_2$ contracts in the loop leg while the other contracts in an external leg. The $\phi$ integral is no longer trivial in this case, and this computation becomes intractable in dim reg. 

Let us consider $\mathcal{Q}$ for such cases. For e.g., Eqn.~\eqref{Eqn:complicated_vertex} in cutoff becomes,
\begin{align}
    \mathcal{Q}=\frac{ \omega_2 p_1 \big( \cos \theta \cos \theta_k + \sin \theta \sin \theta_k \cos \phi \big) + p_1^2}{\omega_2^2 + p_1^2 +2 \omega_2 p_1 \big( \cos \theta \cos \theta_k + \sin \theta \sin \theta_k \cos \phi \big)}\,. \label{Eqn:complicated_vertex_cutoff}
\end{align}
Importantly, one can always express other vertex factors in this form by considering the vector $\vec{k}_i$ in the $\hat{x}-\hat{s}$ plane, where $\hat{x}$ is the axis with respect to which $\phi$ is measured. This makes the calculation of $\int d\phi$ straightforward using the following relation,
\begin{align}
  \int_0^{2\pi} dy  \frac{\alpha +\beta  \cos (y)}{\gamma +\kappa  \cos (y)} = \bigg[ \frac{2 (\beta  \gamma -\alpha  \kappa ) \tanh ^{-1}\left(\frac{(\gamma -\kappa ) \tan \left(\frac{y}{2}\right)}{\sqrt{\kappa ^2-\gamma ^2}}\right)}{\kappa  \sqrt{\kappa ^2-\gamma ^2}}+\frac{\beta  y}{\kappa } \bigg]_0^{2\pi}=2\pi \beta/\kappa \,. \label{Eqn:phi_integral}
\end{align}
Comparing the above with $\mathcal{Q}$, we find $\alpha=p_1^2+\omega_2p_1 \cos \theta \cos \theta_k$, $\kappa=2\omega_2p_1 \sin \theta \sin \theta_k =2\beta$ and $\gamma=\omega_2^2+p_1^2+2\omega_2p_1 \cos \theta \cos \theta_k$ with $y=\phi$. Hence evaluating $\phi$ integral simply gives a factor of $\pi$ and Eqn.~\eqref{Eqn:cutoff_vertex_complicated} becomes,
\begin{align}
    &= \int_0^{\Lambda a(\tau_f)} dp_1 ~p_1^2 Q(p_1,p_2,\{\omega_i\},\{\tau_l\}) \int d\theta ~\sin \theta ~\times \pi \nonumber \\
    &=\pi \int_0^{\Lambda a(\tau_f)} dp_1 \int_{|p_1-s|}^{p_1 +s} dp_2 \frac{p_1 p_2}{s} Q(p_1,p_2,\{\omega_i\},\{\tau_l\}) \,, \label{Eqn:cutoff_loop_integrals}
\end{align}
and we see that the complicated looking $\phi$ integral in Eqn.~\eqref{Eqn:phi_integral} does not produce any $\theta$ dependence. The equality in second line follows from changing variables $\theta \rightarrow p_2$. Now, some of the terms produced in the integral regime $\int_s^{\Lambda a(\tau_f)} \int_{p_1-s}^{p_1+s} dp_1$ are proportional to $\text{exp}\bigg({2\frac{\Lambda}{H} \bigg(\frac{\tau_f}{\tau_l}-1 \bigg)}\bigg)$ (where $\tau_f>\tau_l$). These terms are dropped when computing the time integrals, since these are exponentially suppressed throughout the integration regime and hence will only produce non-log finite terms \cite{Senatore:2009cf}. Finally, evaluation of the momentum integrals produces factors of $\tau_1-\tau_2$ in the denominator. Hence the time integrals are also regulated with the same regulator $\Lambda$ as,
\begin{align}
    \int_{-\infty}^0 d\tau_1\int_{-\infty}^{\tau_1} d\tau_2  \rightarrow   \int_{-\infty}^0 d\tau_1 \int_{-\infty}^{\tau_1 \left(1+ \frac{H}{\Lambda} \right)} d\tau_2 \,.
\end{align}

\section{Results at 1-loop}
\label{Sec:Explicit_Comp_Bispectrum}
In this section, we follow the scheme outlined in Sec.~\ref{Sec:Computational_Scheme} to compute the bispectrum at 1-loop (Fig.~\ref{Fig:Bispectrum}) with cubic and quartic interactions given by $i g_3~\dot{\varphi}^2 \varphi$ and $i g_4 ~\dot{\varphi}^2 \partial^{-2}\partial_i\left(\dot{\varphi}\partial_i\varphi \right)$, taken from the cubic and quartic action Eqns.~\eqref{Eqn:cubic_action} and \eqref{Eqn:quartic_action} and respectively.

\subsection{List of all contractions}
Let us refer to the fields in the interactions as $\varphi_i$, where $i=1,...,7$ to clearly refer to the contractions. Hence, the interactions are $ig_4~\dot{\varphi}_1 \dot{\varphi}_2 \partial^{-2} \partial_i \left( \dot{\varphi}_3 \partial_i \varphi_4\right)$ and $ig_3~ \dot{\varphi}_5 \dot{\varphi}_6 \varphi_7$. The possible contractions for the quartic coupling are given as follows,
\begin{enumerate}
    \item When both the fields being acted upon by the inverse laplacian operator, i.e. $\varphi_3$ and $\varphi_4$, contract in the loop legs, with $\varphi_1$ and $\varphi_2$ contracts in the external legs. 
    \item When the fields not being acted upon by the inverse laplacian operator, i.e. $\varphi_1$ and $\varphi_2$, contract in the loop legs.
    \item When $\varphi_3$ contracts with an external leg, and $\varphi_4$ contracts with a loop leg.
    \item When $\varphi_4$ contracts with an external leg, and $\varphi_3$ contracts with a loop leg.
\end{enumerate}
For each of the contractions mentioned above, we can have the following two cases with the cubic interaction, 
    \begin{enumerate}
        \item The field with no time derivative, i.e. $\varphi_7$,  contracts in the external leg. 
        \item $\varphi_7$ contracts in one of the internal legs.
    \end{enumerate}
    
\begin{figure}[hbt!]
\centering
\includegraphics[scale=0.7]{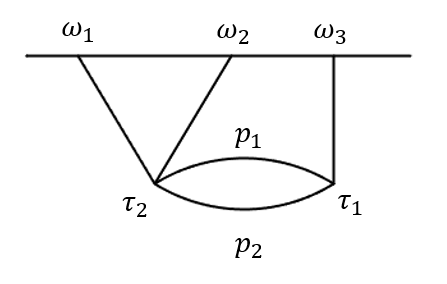}
\caption{Bispectrum at 1-loop computed with one insertion of a cubic and a quartic vertex each.}
\label{Fig:Bispectrum}
\end{figure}

    We will refer to a diagram with Contraction~($i$) of quartic coupling with Contraction~($j$) of cubic coupling as Diagram~$ij$. For example, Diagram~$12$ refers to the diagram in Fig.~\ref{Fig:Bispectrum} where fields being acted upon by inverse laplacian operator in the quartic interaction contract in the loop legs (thus we have $\varphi_1(\omega_1,\vec{k}_1)$, $\varphi_2(\omega_2,\vec{k}_2)$, $\varphi_3(p_1)$ and $\varphi_4(p_2)$), while the field without a time derivative in the cubic interaction contracts in one of the loop legs (hence, we have $\varphi_5(\omega_3,\vec{k}_3)$, $\varphi_6(p_1)$ and $\varphi_7(p_1)$).

    \subsection{Results of explicit computations}
    Below we give the renormalised bispectrum for the Diagrams listed above. We denote the magnitude of the momentum exchanged between the loop sites as $s=|\vec{p}_1+\vec{p}_2|$, which is equal to $k_3$ for this diagram once onshell condition is imposed. Note that we ignore some overall factors of Hubble and external momenta, as well as numerical constants, since we are interested in probing the analytic structure of these correlators. 
    \paragraph{Diagram 11 :} The vertex factor for this diagram is given by,
    \begin{align}
        \frac{\vec{p}_2\cdot \big(\vec{p}_1+\vec{p}_2 \big)}{|\vec{p}_1+\vec{p}_2|^2} = \frac{\vec{p}_2 \cdot \vec{s}}{s^2} = \frac{s^2-p_1^2+p_2^2}{s^2}\,, \label{Eqn:vertex_diagram_11}
    \end{align}
    which is quite simple, and hence can be computed both in cutoff as well as dimensional regularization. 

    In dimensional regularization, considering $3$-d modes and measure, we get,
    \begin{align}
        D_1\sim &\frac{-8 \omega _3^2 \left(s^2+5 \omega _3^2\right)+25 \omega _{12} \omega _3 \left(\omega _3^2-s^2\right)+5 \omega _{12}^2 \left(\omega _3^2-s^2\right)}{\omega_T^5} \bigg(\frac{1}{\delta} -\frac{3}{2} \log \mu \nonumber \\
        &+\frac{1}{2} \log \big[\left(s+\omega _3\right) \left(s+\omega _{12}\right)\big] \bigg) \,,
    \end{align}
    And using Eqn.~\eqref{Eqn:D_2}, the $\mathcal{O}(\delta)$ correction from modes and measure is,
    \begin{align}
        D_2=\frac{3}{2} \delta D_1 \log\big(\frac{H}{\omega_T}\big) +\text{NLf} \,, \label{Eqn:D_2_Diagram_11}
    \end{align}
    since there are $7$ internal modes and $2$ vertex factors, hence $\sum_i n_i=\frac{7}{2}-2=\frac{3}{2}$. Adding $D_1$ and $D_2$, we find \textit{``unphysical"} logarithm of $\omega_T$, which is taken care of once the contact-contribution from counter term is added.  This can be seen without explicitly renormalizing the theory. Generally, the counter-terms are local operators of dimension $8$ (since $[g_3]=-1$ and $[g_4]=-3$), having the following structure,
\begin{align}
    \mathcal{O}^{\text{CT}}\sim \frac{1}{\delta}\int \mu^{-\delta/2} a^{-3+\delta} \partial^{5}\left(\varphi^{3}\right)\,,
\end{align}
where $\partial^{5}$ are 5 space/time derivatives respecting rotational invariance\footnote{Since the kinematic dependence in the coefficient of the divergent term of the loop correlator comes from inverse laplacian operators, it is interesting to ask whether the counter terms might require some non local operators to produce the correct kinematics.}. The contact 3-pt diagram computed with $d$-dimensional modes, renormalizing the divergence in the 1-loop correlator, has $3$ internal modes and $1$ vertex. Thus counter term contribution $D_{\text{c.t.}}$ is given by,
    \begin{align}
        D_{\text{c.t.}}=D_{1,\text{c.t.}}+D_{2,\text{c.t.}} \propto -\bigg(\frac{1}{\delta}-\frac{1}{2}\log \mu +\frac{1}{2}\log \big(\frac{H}{\omega_T} \big)\bigg) \,, \label{Eqn:Diagram_11_ct}
    \end{align}
    where $D_{(1,2),\text{c.t.}}$ denote contact diagram contributions coming from $3$-d and $\mathcal{O}(\delta)$ parts of the modes/measure. The first two terms in Eqn.~\eqref{Eqn:Diagram_11_ct} are from $D_{1,\text{c.t.}}$, while the last term is from $D_{2,\text{c.t.}}$ ($\sum_i n_i = \frac{3}{2}-1=\frac{1}{2}$ coming from $3$ internal modes and $1$ vertex). Adding $D_{\text{c.t}}$ to $D_1+D_2$, we get $D_{\text{renorm.}}=D_{\text{c.t.}}+D_1+D_2$ as,
    \begin{align}
        &D_{\text{renorm.}}\sim \frac{1}{\omega_T^5} \bigg(-8 \omega _3^2 \left(s^2+5 \omega _3^2\right)+25 \omega _{12} \omega _3 \left(\omega _3^2-s^2\right)+5 \omega _{12}^2 \left(\omega _3^2-s^2\right)\bigg) \nonumber \\
        & \bigg(\log \frac{H}{\mu}+\frac{1}{2}\log \big[\frac{\left(s+\omega _3\right)}{\omega_T} \big]+\frac{1}{2}\log \big[\frac{\left(s+\omega _{12}\right)}{\omega_T}\big] \bigg) +\text{NLf} \,,
    \end{align}
    
As a cross check, we calculate this diagram in cutoff regularisation to find that the unrenormalised\footnote{The counter term contributions in cutoff regularization are tree level diagrams computed with $3$-d modes, hence no branch cuts are produced.} correlator features $\log \big(\Lambda/H \big)$ divergence, whose coefficient matches the coefficient of $\log \big(\mu/H\big)$ term in dim reg. The logarithms and their coefficients also match with the above equation.

\paragraph{Diagram 12}
The vertex factor of this diagram is the same as Eqn.~\eqref{Eqn:vertex_diagram_11} and hence this diagram can be computed both in cutoff as well as dimensional regularization. Considering $3$-d modes and measure, we get,
 \begin{align}
        D_1\sim\frac{1}{\omega_T^5} \bigg(8 {\omega_{12}}^2+5 \omega_3^2+31\omega_{12} \omega_3+6 s^2\bigg)\bigg(\frac{1}{\delta} 
        &+\frac{1}{2} \log \big[\left(s+\omega _3\right) \left(s+\omega _{12}\right)\big] -\frac{3}{2} \log \mu \bigg) \,,
    \end{align}
    After following the same reasoning as in Diagram 11 , the renormalized answer $D_{\text{renorm.}}=D_{\text{c.t.}}+D_1+D_2$ turns out to be,
    \begin{align}
        &D_{\text{renorm.}}\sim \frac{1}{\omega_T^5} \bigg(8 {\omega_{12}}^2+5 \omega_3^2+31\omega_{12} \omega_3+6 s^2\bigg)\bigg(2\log \left(\frac{H}{\mu} \right) +\log \left(\frac{\omega_{12}+s}{\omega_T}\right) \nonumber \\
        &+\log \left(\frac{\omega_3+s}{\omega_T}\right)\bigg) +\text{NLf}\,.
    \end{align}
   
     Performing this computation in cutoff regularisation, the unrenormalised correlator features $\log \big(\Lambda/H \big)$ divergence, whose coefficient matches the coefficient of $\log \big(\mu/H\big)$ in dim reg. The logarithms produced in cutoff regularisation, as well as the coefficient, match with the above equation.
     
    \paragraph{Diagram 21}
    The vertex factor of this diagram is given by,
    \begin{align}
        \frac{\vec{k}_2\cdot(\vec{k}_1+\vec{k}_2)}{|\vec{k}+\vec{k}_2|^2} = \frac{\vec{k}_2\cdot \vec{k}_3}{s^2} \,, \label{Eqn:vertex_diagram_21}
    \end{align}
    which is simple enough for the diagram to be computable both in cutoff as well as dimensional regularization. Computing $D_{\text{renorm.}}$, we get,
  \begin{align}
        &D_\text{renorm.}=\frac{\vec{k}_2 \cdot\vec{k}_3}{s^2\omega _T^5} \bigg\{ -\log \left(\frac{H}{\mu} \right) \bigg( 8 s^2 \omega _3 \left(5 \omega _2+\omega _3\right) \left(3 s^2-5 \omega _3^2\right)+8 \omega _2 \omega _{12} \big(10 s^2 \omega _3^2+3 s^4 \nonumber \\
        &-45 \omega _3^4\big) +\omega _1 \big(\omega _3^3 \left(450 \omega _{12}^2-20 s^2\right)+10 \omega _3^2 \left(2 s^2 \omega _{12}+45 \omega _{12}^3\right)+15 \omega _3 \left(2 s^4+15 \omega _{12}^4\right)+6 s^4 \omega _{12} \nonumber \\
        & -45 \omega _3^5+135 \omega _{12} \omega _3^4+45 \omega _{12}^5\big) \bigg) +\log \left(\frac{s+\omega _3}{\omega_T}\right) \bigg(\omega _3 \big(4 s^2 \left(5 \omega _2+\omega _3\right) \left(5 \omega _3^2-3 s^2\right)\nonumber \\
        & +5 \omega _1 \left(2 s^2 \omega _3^2-3 s^4+9 \omega _3^4\right)\big)+\left(\omega _1+4 \omega _2\right) \omega _{12} \left(-10 s^2 \omega _3^2-3 s^4+45 \omega _3^4\right)\bigg) + \log \big(\frac{s+\omega _{12}}{\omega_T}\big) \nonumber \\
        &\bigg(-\omega _1^2 \left(10 \omega _3^2 \left(s^2+135 \omega _2^2\right)+3 \left(s^4+75 \omega _2^4\right)+180 \omega _3^4+900 \omega _2 \omega _3^3+900 \omega _2^3 \omega _3\right)\nonumber \\
        &-5 \omega _1 \left(-2 \omega _3^3 \left(s^2-45 \omega _2^2\right)+10 \omega _2 \omega _3^2 \left(s^2+9 \omega _2^2\right)+3 \omega _3 \left(s^4+15 \omega _2^4\right)+3 \omega _2 \left(s^4+3 \omega _2^4\right)\right) \nonumber \\
        &+4 \left(-3 s^4 \omega _3^2+5 s^2 \omega _3^4+5 s^2 \omega _2 \omega _3 \left(5 \omega _3^2-3 s^2\right)+\omega _2^2 \left(-10 s^2 \omega _3^2-3 s^4+45 \omega _3^4\right)\right)-45 \omega _1^6 \nonumber \\
        &-225 \left(\omega _2+\omega _3\right) \omega _1^5-450 \left(\omega _2+\omega _3\right){}^2 \omega _1^4-450 \left(\omega _2+\omega _3\right){}^3 \omega _1^3\bigg)
        \bigg\} +\text{NLf} \,,  \label{Eqn:Diagram_21} 
    \end{align} 

    Performing the same computation in cutoff regularisation, we find that the coefficients of the logarithms ($\log \big(\frac{\omega_3+s}{\omega_T} \big)$ and $\log \big(\frac{\omega_{12}+s}{\omega_T} \big)$) match. The same is also true for coefficient of $\log\big(\Lambda/H \big)$ (in cutoff) and that of $\log\big(\mu/H \big)$ term (in dim reg). 
    
    In cutoff, we further find power law divergences in $\Lambda$. Some of these terms involve late time log divergences as follows,    
    \begin{align}
     \sim  \frac{3 \Lambda  \left(3 H^2 \Lambda +2 H^3+2 H \Lambda ^2+\Lambda ^3\right) \log \left(\tau _0 \omega_T\right)}{128 H^4\omega_1 \omega_2^3 \omega_3^3} \,. \label{Eqn:naive_secular_div}
    \end{align}
    It is interesting to ask whether this divergence is a \textit{true} secular divergence of the theory, by which we mean whether this divergence persists once one renormalizes the correlator and all power law divergences cancel. Naively it does not seem to be true, since the $\log(\tau_0)$ term does not arise in the dimensional regularization scheme, and in the cutoff regularization it arises only as a coefficient in the power law divergence in $\Lambda$. A true secular divergence in the theory should show up regardless of the regularization scheme. Nevertheless it is important to explicitly demonstrate the existence of the counter terms absorbing the divergence in \eqref{Eqn:naive_secular_div}. Since we do not explicitly renormalize the theory in this paper, we leave this task for a future work.

   \paragraph{Diagram 22} 
   The vertex factor of this diagram is the same as Eqn.~\eqref{Eqn:vertex_diagram_21} and remains computable in cutoff as well as dimensional regularization. Computing $D_{\text{renorm.}}$, we get,
  \begin{align}
        & D_\text{renorm.}\sim \frac{\vec{k}_2\cdot \vec{k}_3}{s^2 \omega_T^5} \bigg(\frac{1}{\delta }+\frac{1}{2} \log \big[\frac{\left(s+\omega _{12}\right) \left(s+\omega _3\right)}{\omega _T^2}\big] \bigg) \bigg( \omega_3\big(-2 \left(5 \omega _2+\omega _3\right) \left(2 s^2+3 \omega _3^2\right) \nonumber \\
        & -\omega _1 \left(5 s^2+21 \omega _3^2\right)\big)-\left(\omega _1+4 \omega _2\right) \omega _{12} \left(s^2+15 \omega _3^2\right) \bigg) +\text{NLf} \,.
    \end{align}
    
Performing this computation in cutoff regularisation gives the same logarithms, and the coefficient of $\log \big(\Lambda/H \big)$ matches the coefficient of $\log\big(\mu/H \big)$ in the previous equation.

   \paragraph{Diagram 31} 
 The phase space integral becomes too complicated to be evaluated in $d-$dimensions, hence instead of dimensional regularization we evaluate this diagram in cutoff regularization. The vertex factor is given by Eqn.~\eqref{Eqn:complicated_vertex_cutoff}. Computing the integrals we get,    
\begin{align}
&D_{\text{unrenorm.}}  
\sim \frac{1}{\omega_T^5} 
\bigg\{
2 \log\left(\frac{\Lambda}{H}\right)
-\log \left(\frac{\omega_{12}+s}{\omega_T}\right)
-\log \left(\frac{\omega_3+s}{\omega_T}\right)
\bigg\}
\bigg\{
-\omega_3^2 \left(3 \omega_{12}^2 + 2 s^2\right) \nonumber \\
&\qquad\qquad\qquad
+5\, \omega_{12}\, \omega_3\, s^2
+\omega_{12}^2 s^2
+15\, \omega_3^4
-6\, \omega_{12}\, \omega_3^3
\bigg\} +\text{NLf} \,.
\end{align}

\paragraph{Diagram 32} Once again we will compute this diagram in cutoff regularization only due to the phase space integral being too complicated in $d-$ dimensions. There are two cases for this diagram : when $\varphi_7$ contracts with $\varphi_4$, and when $\varphi_7$ contracts with $\varphi_1$ (or $\varphi_2$).

We find that in the former case, the loop integral is divergent both in UV as well as IR. For diagrams that are divergent only in the UV, it was noticed that without proper regularization, one finds unphysical features in the loop answer \cite{Senatore:2009cf,Bhowmick:2024kld,Bhowmick:2025mxh}. In dS, the correct way to implement cutoff regularization is to consider a fixed physical cutoff $\Lambda$, and hence when computing the loop integrals, the cutoff $a(\tau_e) \Lambda$ is dependent on time ($\tau_e$ is the time being integrated earliest). However, it is not very clear what the analogous appropriate regularization is in the IR (for related discussions, see \cite{Giddings:2010nc,Byrnes:2010yc,Huenupi:2024ksc,Cespedes:2023aal,Senatore:2012nq,Starobinsky:1994bd} and references therein). A detailed analysis involving regularization of IR divergences, such as the one arising in the above case of Diagram 32, is left for a future work.

    For the second case, the integrals evaluate to,

\begin{align}
&D_\text{unrenorm.}
\sim \frac{1}{\omega_T^5}
\bigg\{
\log\left(\frac{\Lambda}{H}\right)
-\frac{1}{2} \log \left(\frac{\omega_3+s}{\omega_T}\right)
-\frac{1}{2} \log \left(\frac{\omega_{12}+s}{\omega_T}\right)
\bigg\} \big(3\, \omega_{12} \left(\omega_{12}+3 \omega_3\right) \nonumber \\
& + 2 s^2\big) +\text{NLf} \,.
\end{align}

\paragraph{Diagram 41 }
The vertex factor of this diagram is given by,
\begin{align}
    \frac{\vec{k}_2\cdot\big(\vec{k}_2 + \vec{p}_1 \big)}{|\vec{k}_2+\vec{p}_1|^2} \xrightarrow{\text{off-shell}} \frac{ \omega_2 p_1 \big( \cos \theta \cos \theta_k + \sin \theta \sin \theta_k \cos \phi \big) + \omega_2^2}{\omega_2^2 + p_1^2 +2 \omega_2 p_1 \big( \cos \theta \cos \theta_k + \sin \theta \sin \theta_k \cos \phi \big)}\,. \label{Eqn:vertex_diagram_41}
\end{align}

This involves non-trivial angular dependencies, hence is only computed in cutoff regularization. This dependence is handled exactly as in Eqn.~\eqref{Eqn:complicated_vertex_cutoff}, and the $\phi$ integral does not introduce any $\theta$ dependence. Evaluating the integrals we get,

\begin{align}
&D_\text{unrenorm.}
\sim \frac{1}{\omega_T^{5}} \bigg\{ \log \left(\frac{H \left(s+\omega _3\right)}{\Lambda~ \omega _T}\right) \bigg(\omega _3 \big(4 s^2 \left(5 \omega _2+\omega _3\right) \left(3 s^2-5 \omega _3^2\right)+5 \omega _1 \big(-2 s^2 \omega _3^2 \nonumber \\
&+3 s^4-9 \omega _3^4\big)\big)+\left(\omega _1+4 \omega _2\right) \omega _{12} \left(10 s^2 \omega _3^2+3 s^4-45 \omega _3^4\right) \bigg) +\log \left(\frac{H \left(s+\omega _{12}\right)}{\Lambda  \omega _T}\right) \bigg(\omega _1^2 \nonumber \\
&\big(10 \omega _3^2 \big(s^2 +135 \omega _2^2\big) +3 \left(s^4+75 \omega _2^4\right)+180 \omega _3^4+900 \omega _2 \omega _3^3+900 \omega _2^3 \omega _3\big)+5 \omega _1 \big(-2 \omega _3^3 \nonumber \\
&\left(s^2-45 \omega _2^2\right)+10 \omega _2 \omega _3^2 \left(s^2+9 \omega _2^2\right)  +3 \omega _3 \left(s^4+15 \omega _2^4\right)+3 \omega _2 \left(s^4+3 \omega _2^4\right)\big)+4 \big(3 s^4 \omega _3^2 \nonumber \\
&-5 s^2 \omega _3^4+5 \omega _2 \left(3 s^4 \omega _3-5 s^2 \omega _3^3\right)+\omega _2^2 \big(10 s^2 \omega _3^2 +3 s^4-45 \omega _3^4\big)\big)+45 \omega _1^6+225 \omega _{23} \omega _1^5 \nonumber \\
&+450 \omega _{23}^2 \omega _1^4 +450 \omega _{23}^3 \omega _1^3 \bigg) \bigg\} +\text{NLf} \,.
\end{align}

\paragraph{Diagram 42 :} The vertex factor of this diagram is the same as in Eqn.~\eqref{Eqn:vertex_diagram_41}. Computing the diagram in cutoff regularization, we get, 

\begin{align}
& D_\text{unrenorm.}
\sim \frac{1}{\omega_T^{5}}
\bigg\{
2 \log\left(\frac{\Lambda}{H}\right)
- \log\left(\frac{\omega_{12}+s}{\omega_T}\right)
- \log\left(\frac{\omega_3+s}{\omega_T}\right)
\bigg\} \bigg(
\omega_3^{2} \big(15\, \omega_{12} (\omega_1 + 4 \omega_2) \nonumber \\
&
 + 4 s^{2}\big)
+ 5 (\omega_1 + 4 \omega_2)\, \omega_3\, s^{2}
+ \omega_{12} (\omega_1 + 4 \omega_2)\, s^{2}
+ 6\, \omega_3^{4}
+ 3 (7 \omega_1 + 10 \omega_2)\, \omega_3^{3}
\bigg) +\text{NLf} \,.
\end{align}

\paragraph{Singularities of the bispectrum with non-local interactions :} The bispectrum in Fig.~\ref{Fig:Bispectrum} has no \textit{left-, right-subgraphs} (See Appendix.~\ref{Appendix:Diagrammatic_Rules} for definition of the \textit{subgraphs} and summary of the diagrammatic rules to extract singularity structure). The \textit{loop-subgraph} of this diagram has external energy injection of $\mathcal{S}_L=\omega_{12}$ and $\mathcal{S}_R=\omega_3$ through the left and right loop site respectively. The diagrammatic rules tell us that evaluation of this diagram should yield $\log (s+\omega_{12})$ and $\log (s+\omega_3)$.

In all the previously computed results we find,
\begin{itemize}
    \item Logarithms of $s+\omega_3$ and $s+\omega_{12}$. This was precisely expected from the diagrammatric rules. This verifies that the diagrammatic rules hold for the bispectrum computed with the non local interactions in this work.
    \item Dilatation invariant logarithms, where the denominator of the log argument is the total energy $\omega_T$. It has been previously discussed that this total energy branch cut is a feature of dS correlators (computed with polynomial or derivative interactions), and we find this remains true for non-local interactions as well.
 \end{itemize}

The underlying theory considered in this work is local, and only after eliminating the constraints do these non-local terms appear. It may seem that there is a connection between locality of the theory and the diagrammatic rules. The true signature of locality is reflected when every interaction term is included (i.e. locality$= \sum $non-local terms), however we find the diagrammatic rules hold in every single contraction for a given interaction term. If additional branch cuts were generated as a consequence of non-locality, their cancellation would require contributions from the other interaction terms. However, no such additional branch cut is observed in any of the computations. Consequently, we find no conclusive evidence for connection between the diagrammatic rules and locality of the theory.

\section{Other contributions to the Bispectrum with non-local interactions}
\label{Sec:Complicated_Interactions}

In this section, we inspect the structure of the bispectrum with other complicated interactions involving larger number of inverse laplacians.

First, we consider a double ($\partial^{-2}\partial_i (\dot\varphi \partial_i \varphi) 
\partial^{-2} \partial_j (\dot\varphi \partial_j \varphi) $ from Eqn.~\eqref{Eqn:quartic_action}) and a single ($\dot{\varphi} \partial_i \varphi \partial^{-2}\partial_i \dot\varphi$ from Eqn.~\eqref{Eqn:cubic_action}) inverse laplacian operator in the quartic and cubic interaction of the bispectrum (Fig.~\ref{Fig:Bispectrum}). We will only compute one contraction, since the purpose of this computation is not to express the full bispectrum with all interactions at quartic and cubic order, but rather to investigate what happens to the analytic structure in the presence of inverse laplacian operators, which is also why we choose to work with the interactions having the largest number of such operators acting on the fields.

The contraction we compute is described by the following :
\begin{enumerate}
    \item Let us denote the fields in the quartic vertex as $\partial^{-2}\partial_i \bigg(\dot\varphi(\vec{k}_1) \partial_i \varphi(\vec{k}_2)\bigg) 
\partial^{-2} \partial_j \bigg(\dot\varphi(\vec{p}_1) \partial_j \varphi(\vec{p}_2)\bigg) $. The first inverse laplacian operator generates a vertex factor of $\frac{\vec{k}_2\cdot\big(\vec{k}_1+\vec{k}_2\big)}{|\big(\vec{k}_1+\vec{k}_2\big)|^2}$, and the second gives a factor of $\frac{\vec{p}_2\cdot \big(\vec{p}_1+\vec{p}_2 \big)}{|\big(\vec{p}_1+\vec{p}_2\big)|^2} = \frac{s^2+p_2^2-p_1^2}{2s^2}$.
\item We denote the fields in the cubic vertex as $\dot{\varphi}(\vec{p}_2) \partial_i \varphi (\vec{k}_3)\partial^{-2}\partial_i \dot\varphi(\vec{p}_1)$. This inverse laplacian gives a vertex contribution of $\frac{\vec{k}_3\cdot\vec{p}_1}{p_1^2}=\frac{s^2+p_1^2-p_2^2}{2p_1^2}$. 
\end{enumerate}

We evaluate this diagram in dimensional regularization following the scheme outlined in Sec.~\ref{Sec:Computational_Scheme}. As usual, $D_1+D_2$ features unphysical $\log(\frac{\omega_T}{\mu})$, which are removed once contribution from counter terms are added, provided the counter terms renormalize the $1/\delta$ divergence of the correlator, and we get the full renormalized correlator $D_{\text{renorm.}}=D_1+D_2+D_{\text{c.t.}}$ as,

\begin{align}
&D_\text{renorm.} 
\sim \frac{\;\vec{k_2}.\vec{k_3}}{s^2 \omega_T^4}
\bigg[\log \left(\frac{H}{\mu }\right)  \bigg\{\omega _3^2 \big(\omega _1 \left(18 s^2-\omega _2^2+\omega _3^2\right)+36 s^2 \omega _2  +9 s^2 \omega _3 -5 \omega _1^3 \nonumber \\
&-8 \omega _2 \omega _1^2+2 \omega _2^3\big)-\omega _{12} \omega _3 \left(4 \left(\omega _1 \left(\omega _2^2-3 s^2\right)-6 s^2 \omega _2+\omega _1^3+2 \omega _2 \omega _1^2\right)+\left(\omega _1-5 \omega _2\right) \omega _3^2\right) \nonumber \\
& -\omega _{12}^2 \left(\omega _1 \left(\omega _2^2-3 s^2\right)-6 s^2 \omega _2+\omega _1^3+2 \omega _2 \omega _1^2\right)\bigg\} +\frac{1}{2} \log \left(\frac{s+\omega _{12}}{\omega _T}\right) \bigg\{ \omega _3^2 \big(\omega _1 \left(18 s^2-7 \omega _2^2\right)
\nonumber \\
& +36 s^2 \omega _2+9 s^2 \omega _3-11 \omega _1^3-5 \left(4 \omega _2+\omega _3\right) \omega _1^2+2 \omega _2^3+5 \omega _2^2 \omega _3\big)+4 \omega _{12} \omega _3 \big(\omega _1 \left(3 s^2-2 \omega _2^2\right) \nonumber \\
& +6 s^2 \omega _2-2 \omega _1^3-4 \omega _2 \omega _1^2\big)+\omega _{12}^2 \left(\omega _1 \left(3 s^2-2 \omega _2^2\right)+6 s^2 \omega _2-2 \omega _1^3-4 \omega _2 \omega _1^2\right) \bigg\} \nonumber \\
&+ \frac{1}{2} \log \left(\frac{s+\omega _3}{\omega _T}\right) \bigg\{ \omega _3^2 \left(9 s^2 \left(4 \omega _2+\omega _3\right)+2 \omega _1 \left(9 s^2+\omega _3^2\right)\right)+\omega _{12} \omega _3 \big(12 s^2 \left(\omega _1+2 \omega _2\right) \nonumber \\
& +\left(3 \omega _1+5 \omega _2\right) \omega _3^2\big)+\left(\omega _1+2 \omega _2\right) \omega _{12}^2 \left(3 s^2+\omega _3^2\right) \bigg\} \bigg] +\text{NLf} \,.
\end{align}

Interestingly, the singularity structure of the bispectrum computed with non-local interactions once again follows the diagrammatic rules. The arguments in the logarithms are $(s+\omega_{12})$ and $(s+\omega_3)$, which are exactly what one expects from \textit{loop-subgraph} contributions, with the total energy branch cut combining to give scale invariant, dimensionless logarithms. No additional logarithms are produced.

\begin{figure}[hbt!]
\centering
\includegraphics[scale=0.7]{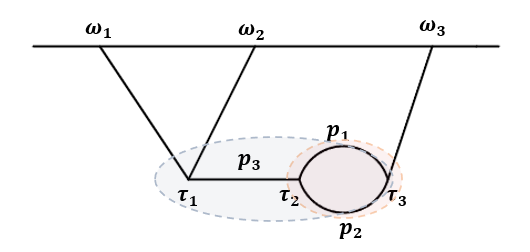}
\caption{The bispectrum at 1-loop computed with three insertions of cubic vertex.}
\label{Fig:3_3v_pt}
\end{figure}

There is yet another contribution to the $1$-loop bispectrum (Fig.~\ref{Fig:3_3v_pt}) obtained by three insertions of cubic operator. As compared to the diagram in Fig.~\ref{Fig:Bispectrum}, this contribution has non trivial subgraphs : the \textit{left-} and the \textit{loop-subgraphs} are shaded in blue and red in Fig.~\ref{Fig:3_3v_pt} respectively. According to the diagrammatic rules, the contribution expected from this subgraph is $\log(\mathcal{S}_L+s)=\log(p_3+s)$ and $\log(\mathcal{S}_R+s)=\log(\omega_3+s)$, where $\mathcal{S}_L=p_3$ and $\mathcal{S}_R=\omega_3$ are the external energy injection into the \textit{loop} subgraph through the left and right loop sites. The \textit{left-}subgraph should contribute a $\log(\mathcal{S}_{L_1}+s)=\log(\omega_{12}+s)$, where $\mathcal{S}_{L_1}$ is the external energy injection into the \textit{left}-subgraph through all vertices except the right loop site. This diagram provides a good premise to investigate what happens to the branch cuts beyond simple graphs such as Fig.~\ref{Fig:Bispectrum} when non-local interactions are involved. 

We consider all three interaction vertices of this diagram to be the cubic interaction $\dot{\varphi} \partial_i \varphi \partial^{-2}\partial_i \dot\varphi$ taken from Eqn.~\ref{Eqn:cubic_action}. With this, we compute the following contraction :
\begin{enumerate}
    \item At $\tau_1$, the cubic vertex is $\dot{\varphi}(\vec{k}_2) \partial_i\varphi(\vec{p}_3) \partial^{-2}\partial_i\dot{\varphi}(\vec{k}_1)$, which generates a vertex factor of $\frac{\vec{p}_3 \cdot \vec{k}_1}{|\vec{k}_1|^2}$.
    \item At $\tau_2$, the cubic vertex is $\dot{\varphi}(\vec{p}_2) \partial_i\varphi(\vec{p}_3) \partial^{-2}\partial_i\dot{\varphi}(\vec{p}_1)$, which generates a vertex factor of $\frac{\vec{p}_3 \cdot \vec{p}_1}{|\vec{p}_1|^2}=\frac{s^2-p_2^2+p_1^2}{2p_1^2}$.
     \item At $\tau_3$, the cubic vertex is $\dot{\varphi}(\vec{p}_1) \partial_i\varphi(\vec{k}_3) \partial^{-2}\partial_i\dot{\varphi}(\vec{p}_2)$, which generates a vertex factor of $\frac{\vec{k}_3 \cdot \vec{p}_2}{|\vec{p}_2|^2}=\frac{s^2+p_2^2-p_1^2}{2p_2^2}$.
\end{enumerate}

Again, the unphysical $\log(\frac{\omega_T}{\mu})$ leftover from $D_1+D_2$ are removed once contribution from counter terms $D_{\text{c.t.}}$ is added (if the counter terms cancel the $1/\delta$ divergence of the loop), and the final renormalized correlator is given by,
\begin{align}
    &D_{\text{renorm.}} \sim \frac{s\;\vec{k_1}.\vec{k_3}}{\omega_T^4 \left(p_3+\omega _3\right){}^3 \left(p_3+\omega _{12}\right){}^2} \bigg[\bigg(\log \left(\frac{H}{\mu }\right)+\frac{1}{2} \log \bigg(\frac{s+\omega _3}{\omega_T}\bigg)  \bigg)\bigg\{ -p_3^2 \big(18 s^2 \omega _3 \nonumber \\
    &+\omega _{12} \left(3 s^2+\omega _3^2\right)\big) \omega _T^4 -3 p_3 s^2 \omega _3 \left(2 \omega _3+3 \omega _{12}\right) \omega _T^4+2 p_3^6 \left(\omega _{12} \left(3 s^2+\omega _3^2\right)-2 \omega _3 \left(\omega _3^2-6 s^2\right)\right) \nonumber \\
    & +2 p_3^5 \left(3 \omega _3+2 \omega _{12}\right) \left(\omega _{12} \left(3 s^2+\omega _3^2\right)-2 \omega _3 \left(\omega _3^2-6 s^2\right)\right)-6 p_3^4 \omega _3 \left(2 \omega _3+3 \omega _{12}\right) \nonumber \\
    &\left(-5 s^2 \omega _3-2 s^2 \omega _{12}+\omega _3^3\right)-2 p_3^3 \big(3 s^2 \omega _3^4+\omega _{12} \omega _3^3 \left(7 \omega _3^2-3 s^2\right)+6 \omega _{12}^2 \omega _3^2 \left(2 s^2+\omega _3^2\right) \nonumber \\
    & +4 \omega _{12}^3 \omega _3 \left(3 s^2+\omega _3^2\right)+\omega _{12}^4 \left(3 s^2+\omega _3^2\right)+\omega _3^6\big)-3 s^2 \omega _3^2 \omega _{12} \omega _T^4 \bigg\} + \frac{\log \big(\frac{s+\omega _{12}}{\omega_T}\big)}{\left(p_3-\omega _{12}\right){}^2}  \nonumber \\
    & \bigg\{p_3^3 \left(p_3+\omega _3\right){}^3 \big(-2 \omega _3^3 \left(p_3^2+3 s^2\right)+\omega _{12} \omega _3^2 \left(p_3^2-3 \left(8 s^2+\omega _{12}^2\right)\right)+12 s^2 \omega _3 \left(p_3^2-3 \omega _{12}^2\right) \nonumber \\
    & +3 s^2 \omega _{12} \left(p_3^2-3 \omega _{12}^2\right)\big) \bigg\}-\frac{\log \left(\frac{p_3+s}{\omega_T}\right)}{2 \left(p_3-\omega _{12}\right){}^2} \bigg\{ \omega_T^4 \left(\omega _{12}-2 p_3\right) \left(p_3+\omega _{12}\right){}^2\nonumber \\
    &\left(9 p_3 s^2 \omega _3+p_3^2 \left(3 s^2+\omega _3^2\right)+3 s^2 \omega _3^2\right) \bigg\}  \bigg]  +\text{NLf}
    \,, \label{Eqn:3_v_dS_full}
\end{align}
 Once again, the logarithms are correctly predicted by the diagrammatic rules , which combines with the total energy branch cut produced from $\mathcal{O}(\delta)$ corrections of modes, to give a scale invariant correlator. The apparent folded poles at $p_3=\omega_{12}$ cancels from the non-log finite terms ($\text{NLf}$). 

\section{Polology}
\label{Sec:Polology}
In this section, we study the connection between the flat space scattering amplitudes and the cosmological correlators corresponding to the same process. It was previously noticed that the leading singularity of cosmological correlators in the limit $k_T\rightarrow0$ is related to the flat space scattering amplitude corresponding to the same process \cite{Maldacena:2011nz,Raju:2012zr,Arkani-Hamed:2017fdk,Arkani-Hamed:2018kmz,Benincasa:2018ssx,Baumann:2020dch,Goodhew:2020hob}. The intuitive way of understanding why such a relation should exist is to note that the exponential $i\epsilon$ suppression in the Bunch Davies modes at early time goes away when $k_T\rightarrow0$, hence this limit probes the early time era when the modes are well approximated by the flat space limit. This was proved at tree level (and for loop integrands) for massless scalars in \cite{Goodhew:2020hob}, where the authors developed Feynman rules analogous to those of cosmological correlators, equipped with different propagators for bulk-bulk (exchanges) and bulk-boundary (external particles), as well as vertex factors coming from general (non-Lorentz invariant) interactions involving fields with time and/or spatial derivatives. 

\subsection{Polology at 1-loop}
Below we review this connection and show that this relation is also satisfied beyond tree level diagrams, at the level of the full $1$-loop correlator. We then demonstrate this with explicit examples of loop correlators, computed with derivative as well as non-local interactions.

Consider interactions of the type $\partial_i^{r_1}\partial_\tau^{s_1} \phi_1...\partial_i^{r_n}\partial_\tau^{s_n} \phi_n$. Here, every field $\phi_k$ is being acted upon by $r_k$ spatial derivatives/inverse laplacians ($r_k=1$ for $\partial_i$, $r_k=-2$ for $\partial^{-2}$) and $s_k$ time derivatives ($\partial_\tau$). An important thing to note here is that since the same interactions are being used to compute scattering amplitudes as well as the dS correlators, the spatial derivatives will generate the same factors of dot products, since they act on $e^{i \vec{k}\cdot \vec{x}}$ in both cases. As we argue later, this allows us to extend our arguments to interactions involving inverse laplacian operators. Thus the spatial derivatives will cause no trouble; we will plug back their contribution when actually evaluating the correlators/scattering amplitudes, however we can ignore these factors for now. 

The time derivatives, on the other hand, need to be considered carefully. They act on the flat space modes $f_k(t)=\frac{1}{\sqrt{2k}} e^{ikt}$ and the ($3$-d) dS modes $g_k(\tau)=\frac{H}{\sqrt{2k^3}}(1-ik\tau)e^{ik\tau}$ as,
\begin{align}
    \partial_t^s f_k(t)=F_s(k) f_k(t), \hspace{10pt} \partial_\tau^s g_k(\tau)= F_s(k) \frac{H}{\sqrt{2k^3}}(1-s-ik\tau)e^{ik\tau} \,,
\end{align}
where $F_s(k)$ is a factor of $k$ raised to a power depending on $s$.

With this, the Feynman rules (re)formulated in \cite{Goodhew:2020hob} are summarized as,
\begin{enumerate}
    \item For every external particle connected to an interaction vertex at $t$ ($\tau$), in the scattering amplitudes we have,
    \begin{align}
        G_{\text{B,flat}} (k,t)= F_s(k) e^{ikt} \,,
    \end{align}
     and for the dS correlators we have,
    \begin{align}
        G_{B,+} (k,\tau)=\frac{H^2}{2k^3} (-H\tau)^s F_s(k)(1-s-ik\tau) e^{ik\tau}\,, \label{Eqn:bulk_boundary_dS}
    \end{align}
     where the $+$ in the subscript of the propagator denotes the interaction is inserted below  the $\tau=0$ line, coming from the time ordered part of the in-in correlator. Complex conjugation of this gives $G_{B,-}$ : the propagator corresponding to the interaction insertion above $\tau=0$. The factor of $(-H \tau)^s$ comes from demanding scale invariance in the interactions.
    \item For every exchange between two interaction vertices at $t_1$ ($\tau_1$) and $t_2$ ($\tau_2$), with $s_1$ and $s_2$ number of time derivatives acting on the exchange fields at the two vertices respectively, for the flat space scattering amplitudes we have,
    \begin{align}
        G_{\text{flat}}(p,t_1,t_2)=F_{s_1}(p) F_{s_2}^*(p) \frac{e^{ip(t_1-t_2)}}{2p} \theta(t_2-t_1) + (t_2\leftrightarrow t_1,s_1 \leftrightarrow s_2) \,,
    \end{align}
    and for the dS correlators we have,
    \begin{align}
        G_{++}(p,\tau_1,\tau_2)&= F_{s_1}(p) F_{s_2}^*(p) \frac{H^2}{2p^3} (-H\tau_1)^{s_1} (-H\tau_2)^{s_2} (1-s_1-ip\tau_1) \nonumber \\
        & (1-s_2+ip\tau_2) e^{ip(\tau_1-\tau_2)} \theta(\tau_2-\tau_1) + (\tau_2\leftrightarrow \tau_1,s_1 \leftrightarrow s_2) \,, \label{Eqn:bulk_bulk_dS}
    \end{align}
    where the $+$ in the subscript again has the same implication as stated previously. One can similarly obtain $G_{+-},G_{-+}$ and $G_{--}$, although we will never use the first two. This is because diagrams which have vertices in both upper and lower region of $\tau=0$ will never produce a $k_T$ pole, and hence we will only look at totally (un-)ordered diagrams. 
    \item Every interaction vertex and undetermined (3-) momenta is integrated over. Ofcourse the time integrals for scattering amplitude computations are evaluated all the way from $-\infty$ to $\infty$, whereas for correlators they are evaluated till $\tau_0$ : the time at which these correlators are being computed. 
\end{enumerate}

We will only consider IR finite interactions and plug in $\tau_0=0$. It can be checked that these rules give back the familiar form of scalar propagator in an exchange process with $s_1=s_2=0$ \cite{Goodhew:2020hob}.

Note that, keeping only terms with the highest power in $\tau$, we see that the dS propagators satisfy,
\begin{align}
    &G_{B,+}\xrightarrow{\tau^\text{max}} i G_{\text{B,flat}}(k,\tau) (-1)^{s+1}\frac{H^{2+s}}{2k^2}\tau^{s+1} e^{ik \tau} \nonumber \,, \\
    & G_{++}\xrightarrow{\tau^\text{max}}G_{\text{flat}}(p,\tau_1,\tau_2) (-1)^{s_1+s_2} H^{2+s_1+s_2} \tau_1^{s_1+1} \tau_2^{s_2+1} \,. \label{Eqn:flat_dS_lead_prop}
\end{align}
hence the exchange propagators are related by factors independent of the exchange momenta. For diagrams involving loops, we must use $d$-dimensional modes in Eqn.~\eqref{Eqn:modes_d_dim_expand} and split the computation of the full correlator $D$ into $D_1$ and $D_2$ as outlined in Sec.~\ref{Sec:Computational_Scheme}. Since at $1$-loop, $D_2$ is immediately obtained from divergent part of $D_1$ multiplied to $(-\log \omega_T)$ (see Appendix~\ref{App_D_2}), we will first focus on the flat space limit of $D_1$, and comments on relation to $D_2$ will follow later. Since the propagators in $D_1$ are given by Eqns.~\eqref{Eqn:bulk_boundary_dS} and \eqref{Eqn:bulk_bulk_dS}, the terms in the loop integrand of dS correlators involving highest power of $\tau$, and the corresponding flat space amplitude are related by factors which are independent of loop momenta (Eqn.~\eqref{Eqn:flat_dS_lead_prop}), and hence these factors can be pulled out of the loop integral $D_1$. In the discussion below, we focus only on $D_1$ unless mentioned otherwise.

There is a crucial ingredient in realizing the relationship between flat space scattering amplitude and the leading singular terms of the dS correlator. Consider the form of a nested time integral arising in the correlator,
\begin{align}
    \int_{-\infty}^0 d\tau_1 \int_{-\infty}^{\tau_1} d\tau_2....\int_{-\infty}^{\tau_{n-2}}d\tau_{n-1} \int_{-\infty}^{\tau_{n-1}} d\tau_n ~e^{-a_1\tau_1}...e^{ia_n\tau_n} Q(\omega_i,p_j,\tau) \,,
\end{align}
where $a_1,...,a_n$ are functions of external and exchange momenta, adding up to the total energy $\omega_T$, and $Q$ is some  function of external energies, exchange momenta and $\tau$'s. $Q$ is analytic in $\tau$'s, i.e. all time arguments appear in the numerator, since we work with IR finite interactions. All intermediate integrals would mix arguments of the exponential, leading to $e^{i \omega_T\tau_1}$ in the final time integral, which would produce total energy pole. If we are interested in the most singular term in the limit $\omega_T \rightarrow0$, then we must only bother with the highest power of $\tau_1$ - the vertex to be integrated last. This highest power is produced from integrating (over $d\tau_2$) the term which has the largest power of $\tau_2$, which in turn comes from integrating (over $d\tau_3$) the term having the largest power of $\tau_3$ and so on. This implies, to extract the leading singularity, we only have to keep track of the term with the largest power of $\tau$ every step of the way.

Note that we are interested in establishing the relation between the \textit{off-shell} dS correlator and the \textit{off-shell} scattering amplitude. An off-shell correlator, defined in Sec.~\ref{Sec:Notations}, as well as an off-shell amplitude is obtained when the energy of the external legs $\omega_k$ is treated as being independent of the momenta $\vec{k}$. An off-shell correlator has internal legs on-shell ($\omega_{p_i}=p_i=|\vec{p}_i|$) and the loop integrand develops poles when one or more vertices become energy conserving, while an off-shell scattering amplitude has energy conserved and loop integrand develops poles when one or more internal particles goes on-shell. 

\begin{figure}[hbt!]
\centering
\includegraphics[scale=0.4]{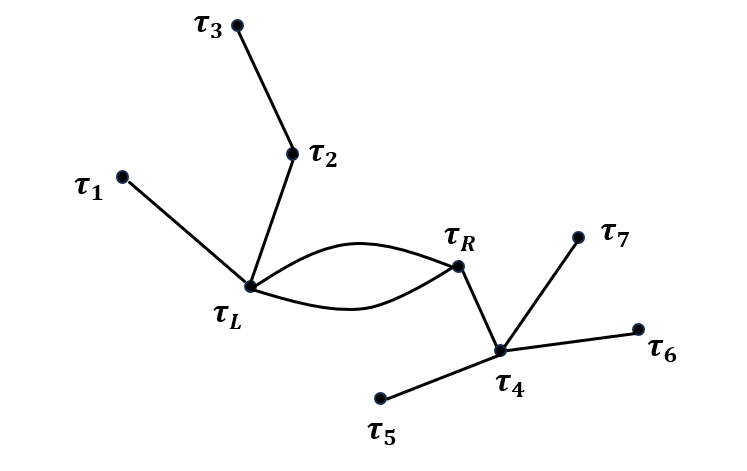}
\caption{A correlator/scattering amplitude at 1-loop and 2 loop sites $\tau_L$ and $\tau_R$. The nodes denote interaction vertices, labeled by $\tau_i$, the exchange/internal legs are marked by lines, and the external lines connected to the interaction vertices are not drawn.}
\label{Fig:1_loop_polology}
\end{figure}

Let us now consider a 1-loop correlator having $2$ loop sites $\tau_L$ and $\tau_R$, with all ordered vertices. We will compute the dS correlator and the flat space scattering amplitude corresponding to the process in Fig.~\ref{Fig:1_loop_polology}, using dimensional regularization, hence the time integrals will be computed first. We denote external energy by $\omega_i$ and exchange as $p_j$, with $p_{l_1}$ and $p_{l_2}$ as the loop momenta. In Fig.~\ref{Fig:1_loop_polology} the nodes denote interaction vertices, labeled by $\tau$ ($t$ for scattering amplitude), exchanges are marked by lines. The interaction vertices can have any number of external lines, which are not drawn in this representation. Since this diagram allows only one closed loop at $\tau_L-\tau_R$, it is possible to find atleast 1 endpoint, i.e. a vertex which is connected to only one other vertex. We denote this vertex by $\tau_{\text{end}}$, and we label its neighbor as $\tau_\text{adj}$. To solve the time integrals, we integrate over $\tau_{\text{end}}$, and move on to the next endpoint vertex in the graph.

First let us consider the case where neither loop site is an endpoint. Considering the flat space scattering amplitude, from the endpoint vertex $V$, (i.e. $t_{\text{end}}$) we have,
\begin{align}
  \mathcal{A}_n : &~\mathcal{F}_V(\vec{k},\vec{p})\int dt_{\text{end}} \bigg(\prod_{i \in v} G_{B,\text{flat}}(\omega_i,t_{\text{end}}) \bigg) G_{\text{flat}}(p,t_{\text{end}},t_{\text{adj}}) \nonumber\\
  &=\mathcal{F}_V(\vec{k},\vec{p}) \int dt_{\text{end}} \bigg(\prod_{i \in v} F_{s_i} (\omega_i) e^{i\omega_i t_{\text{end}}} \bigg) \bigg[ \frac{F_{s_{t_\text{end}}}(p)F^*_{s_{t_\text{adj}}}(p)}{2p} e^{-ip(t_{\text{adj}}-t_{\text{end}})} \theta(t_{\text{adj}}-t_{\text{end}}) \nonumber \\
  &+ \bigg( t_{\text{adj}} \leftrightarrow t_{\text{end}}\bigg) \bigg] 
  \,, \label{Eqn:A_n_t_end}
\end{align}
where $v$ denotes external legs attached to $V$, and $p$ is exchange of $t_{\text{end}}$ with $t_{\text{adj}}$. $s_{\text{end}}$ and $s_{t_\text{adj}}$ are the number of time derivatives acting on the exchange field at $t_\text{end}$ and $t_{\text{adj}}$ respectively.
Although we keep track of the vertex factors $\mathcal{F}_{V}$ coming from the action of $\partial_i$ and $\partial^{-2}$ on the modes, these operators produce the same vertex factors in flat space scattering amplitudes and dS correlators, and hence we need not worry about them for the purpose of this exercise. 

Evaluation of the integral gives \cite{Goodhew:2020hob},
\begin{align}
    \mathcal{A}_n:  \mathcal{F}_V(\vec{k},\vec{p}) \bigg( \prod_{i \in v} F_{s_i} (\omega_i) \bigg) \mathcal{G}(p) e^{i (\sum_{i \in v}\omega_i)t_{\text{adj}}} \,, \label{Eqn:A_n_t_end_ans}
\end{align}
where,
\begin{align}
    \mathcal{G}(p)=\frac{F_{s_{t_\text{end}}}(p)F^*_{s_{t_\text{adj}}}(p)}{2p (p+ \sum_{i \in v} \omega_i)} + \frac{F_{s_{t_\text{adj}}}(p)F^*_{s_{t_\text{end}}}(p)}{2p (p- \sum_{i \in v} \omega_i)}\,.
\end{align}

Similarly, integrating end point vertex $V$ (i.e. $\tau_{\text{end}}$) in the $D_1$ contribution to the dS correlator gives,
\begin{align}
    D_{1_n}: \mathcal{F}_V(\vec{k},\vec{p}) (-H)^{-4+R_V} \int d\tau_{\text{end}} \tau_{\text{end}}^{-4+R_V} \bigg(\prod_{i \in v} G_{B,+}(\omega_i,\tau_{\text{end}}) \bigg) \bigg(G_{++}(p,\tau_\text{end},\tau_{\text{adj}}) \bigg)\,,
\end{align}
where $R_V=\bigg(\sum_{i \in v} r_i\bigg) +r_{p,\tau_{\text{end}}}$ is the total number of spatial derivatives/inverse laplacians acting on the fields at $V$. Note $r_i=1,-2$ for a spatial derivatives and an inverse laplacian operators respectively acting on the $i^{\text{th}}$ field. The factors of $\tau$ involving powers of $R_{V_i}$ arise due to demanding scale invariant interactions in the presence of spatial derivatives. As previously discussed, we will focus on keeping only the terms leading in $\tau$, and using Eqn.~\eqref{Eqn:flat_dS_lead_prop}, we find,
\begin{align}
    D_{1_n}:~& \mathcal{F}_V(\vec{k},\vec{p}) (H)^{-4+R_V} \bigg( \prod_{i \in v} \frac{-i H^2}{2\omega_i} (-H)^{s_i} F_{s_i}(\omega_i) \bigg) (-H)^{s_{p,\tau_\text{end}}+s_{p,\tau_\text{adj}}+2} \tau_{\text{adj}}^{s_{p,\tau_{\text{adj}}}+1} \nonumber \\
    & \int d\tau_{\text{end}} \tau_\text{end}^{-4+R_V+v+(\sum_{i \in v}s_i)+s_{p,\tau_{\text{end}}}+1} e^{i\tau_{\text{end}}(\sum_{i\in v}\omega_i)}~~ \times \nonumber \\
    & \bigg[\frac{F_{s_{\tau_\text{end}}}(p)F^*_{s_{\tau_\text{adj}}}(p)}{2p}e^{ip(\tau_\text{end}-\tau_\text{adj})} \theta(\tau_\text{adj}-\tau_{\text{end}}) +\bigg(\tau_\text{end} \leftrightarrow\tau_\text{adj} \bigg) \bigg] \,,
\end{align}
wherer we have used $(-1)^{(-4+R_V)}=1$ since the total number of spatial derivatives in any interaction vertex must be even due to rotational invariance. This integral has the same structure as the flat space one in Eqn.~\eqref{Eqn:A_n_t_end} modulo some factors, such as extra powers of $\tau_\text{end}$, and the upper limit of the time integral in the $\theta(\tau_\text{end}-\tau_\text{adj})$ piece. The difference in the upper limit is irrelevant here since this limit produces terms subleading in powers of $\tau_{\text{adj}}$. Hence, keeping only the highest power of $\tau_\text{adj}$, evaluation of $\tau_\text{adj}$ gives,
\begin{align}
    D_{1_n}:&~(H)^{-4+R_V+2v+S_V+2} (-H)^{s_{p,\tau_{\text{adj}}}} (i)^{v}(-1)^{S_V+v} \bigg(\prod_{i \in v} \frac{1}{2\omega_i^2}\bigg)\tau_{\text{adj}}^{-4+2+s_{p,\tau_\text{adj}}+R_V+S_V+v} \nonumber \\
    & \times \text{flat space expression Eqn.~\eqref{Eqn:A_n_t_end_ans}} \,, \label{Eqn:t_end_reltn}
\end{align}
 where $S_V=(\sum_{i \in v}s_i)+s_{p,\tau_\text{end}}$ is the total number of temporal derivatives acting on all the fields at $V$. Importantly, note that the flat space answer is recovered due to Eqn.~\eqref{Eqn:flat_dS_lead_prop} only when keeping highest power of $\tau$. If one keeps subleading terms as well, then the factors in Eqn.~\eqref{Eqn:t_end_reltn} become dependent on the exchange $p$. 

Going back to the flat space scattering amplitude computation, we see from Eqn.~\eqref{Eqn:A_n_t_end_ans} that every endpoint vertex integrated out contributes a factor of,
\begin{align}
    \mathcal{F}_V(\vec{k},\vec{p}) \bigg( \prod_{i \in v} F_{s_i}(\omega_i) \bigg) \mathcal{G}(p) e^{i t_\text{adj} (\sum_{i \in v}\omega_i)} \,, \label{Eqn:t_adj_t_end_contr}
\end{align}
where the exponential modifies the integral structure of the neighboring vertex $t_{\text{adj}}$. Now, $t_\text{adj}$ may not be an endpoint. For example, in Fig.~\ref{Fig:1_loop_polology}, there are multiple choices for endpoint. If we start with $t_\text{end}=\tau_3$, then $t_\text{adj}=\tau_2$ becomes the next endpoint when $t_\text{end}$ is integrated over. However, with $t_\text{end}=\tau_2$, $t_\text{adj}=\tau_L$ does not become an endpoint when $\tau_2$ is integrated. The next endpoint can be considered to be $\tau_1$. Importantly, considering different endpoints will only affect the order of integration, however the end result remains the same. This is because the contribution modifying the integral structure of $t_\text{adj}$ is simply additive : it does not matter in which order you consider $\tau_5,\tau_6$ and $\tau_7$ as endpoints, finally one ends up with the same factor for $t_\text{adj}=\tau_4$,
\begin{align}
    \bigg(\prod_{i \in v_5,v_6,v_7} \mathcal{F}_i (\vec{k},\vec{p})  F_{s_i}(\omega_i) \bigg) \bigg(\prod_{\beta}^{\text{ex}} \mathcal{G}(p_\beta)\bigg) e^{i t_{\text{adj}}(\sum_{i \in v_5,v_6,v_7} \omega_i)} \,.
\end{align}

Let us next consider the case when we hit the loop site as the endpoint. Suppose we have $t_\text{end}=\tau_L$. The neighbouring vertex is $t_{\text{adj}}=\tau_R$ with which it exchanges $p_{l_1}$ and $p_{l_2}$, so for the integral over $\tau_L$ we have,
\begin{align}
    \mathcal{A}_n:&~ \mathcal{F}_{\tau_L}(\vec{k},\vec{p}) \int d\tau_L \bigg( \prod_{i \in v_{\tau_L}}G_{B,\text{flat}}(\omega_i,\tau_L) \bigg) G_{\text{flat}}(p_{l_1},\tau_L,\tau_R) G_{\text{flat}}(p_{l_2},\tau_L,\tau_R) \nonumber \\
    &=\mathcal{F}_{\tau_L}(\vec{k},\vec{p}) \bigg(\prod_{i \in v_{\tau_L}} F_{s_i} (\omega_i) \bigg) \bigg\{ \int_{-\infty}^{\tau_R} d\tau_L e^{i(\sum_{i \in v_{\tau_L}}\omega_i)\tau_L} \frac{e^{i(p_{l_1}+p_{l_2})(\tau_L-\tau_R)}}{4p_{l_1}p_{l_2}}\nonumber \\
    &F_{s_{\tau_L}}(p_{l_1})F_{s_{\tau_R}}^*(p_{l_1}) F_{s_{\tau_L}}(p_{l_2}) F_{s_{\tau_R}}^*(p_{l_2})+\int_{\tau_R}^{\infty}d\tau_L e^{i(\sum_{i \in v_{\tau_L}}\omega_i)\tau_L} \frac{e^{i(p_{l_1}+p_{l_2})(\tau_R-\tau_L)}}{4p_{l_1}p_{l_2}} \nonumber \\
    & F^*_{s_{\tau_L}}(p_{l_1})F_{s_{\tau_R}}(p_{l_1}) F^*_{s_{\tau_L}}(p_{l_2}) F_{s_{\tau_R}}(p_{l_2}) \bigg\} \nonumber \\
    & =\mathcal{F}_{\tau_L}(\vec{k},\vec{p})\bigg(\prod_{i \in v_{\tau_L}} F_{s_i} (\omega_i) \bigg) \mathcal{H}(p_{l_1},p_{l_2})  e^{i(\sum_{i \in v_{\tau_L}}\omega_i)\tau_R} \,, \label{Eqn:t_loop_site}
\end{align}
where,
\begin{align}
    \mathcal{H}(p_1,p_2)=\frac{F_{s_{\tau_L}}(p_1)F_{s_{\tau_R}}^*(p_1) F_{s_{\tau_L}}(p_2) F_{s_{\tau_R}}^*(p_2)}{4p_1p_2 (p_1+p_2+\sum_{i \in v_{\tau_L}}\omega_i)}+\frac{F^*_{s_{\tau_L}}(p_1)F_{s_{\tau_R}}(p_1) F^*_{s_{\tau_L}}(p_2) F_{s_{\tau_R}}(p_2)}{4p_1p_2 (p_1+p_2-\sum_{i \in v_{\tau_L}}\omega_i)}\,, \label{Eqn:loop_H}
\end{align}
which is exactly what one finds as the integrand when performing integral over the loop ($3$-)momenta in usual flat space computations with $s_{\tau_L}=s_{\tau_R}=0$ (See Appendix.~\ref{App_Flat_loop}). 

Using Eqn.~\eqref{Eqn:t_loop_site} we see that contribution from a loop site endpoint is the same as that of Eqn.~\eqref{Eqn:t_adj_t_end_contr}, except instead of $\mathcal{G}(p)$, loop site end points will contribute $\mathcal{H}(p_1,p_2)$. The remaining time integrals are solved with the same algorithm until one is left with the final time integral, which is just a contact diagram giving energy conservation. Hence, the (energy momentum conserving delta function stripped) flat space correlator is given by the following integral over loop momentum $p_l$,
\begin{align}
    \mathcal{A}_n=\int d^3\vec{p_l} \bigg(\prod_{\alpha \in \text{vertices}} \mathcal{F}_\alpha (\vec{k},\vec{p})\bigg) \bigg( \prod_{i\in \text{ext legs}}F_{s_i}(\omega_i) \bigg) \bigg(\prod_{\beta}^{\text{ex}} \mathcal{G}(p_\beta) \bigg) \mathcal{H}(p_{l_1},p_{l_2})\,.
\end{align}
Finally, to see how Eqn.~\eqref{Eqn:t_end_reltn} modifies when considering a loop site in a correlator as an endpoint, we have, considering $\tau_L$ as the endpoint and only retaining terms with highest powers of $\tau_L$ (and $\tau_R$) as before,
\begin{align}
    &D_{1_n}:\mathcal{F}_{V_{\tau_L}}(\vec{k},\vec{p}) (H)^{R_V} \bigg( \prod_{i \in v} \frac{-i H^2}{2\omega_i^2} (-H)^{s_i} F_{s_i}(\omega_i) \bigg) (-H)^{s_{p_{l_1},\tau_L}+s_{p_{l_2},\tau_L}+s_{p_{l_1},\tau_R}+s_{p_{l_2},\tau_R}} \nonumber \\
    &\times (\tau_R)^{s_{p_{l_1},\tau_R}+s_{p_{l_2},\tau_R}+2} \int d\tau_L (\tau_L)^{-4+R_V+(\sum_{i\in v} s_i)+v+s_{p_{l_1},\tau_L}+s_{p_{l_2},\tau_L}+2} e^{i(\sum_{i \in v} \omega_i)\tau_L}~\times \nonumber \\
    & \bigg[ \frac{e^{i(p_{l_1}+p_{l_2})(\tau_L-\tau_R)}}{4p_{l_1} p_{l_2}} F_{s_{\tau_L}}(p_{l_1})F_{s_{\tau_R}}^*(p_{l_1}) F_{s_{\tau_L}}(p_{l_2})F_{s_{\tau_R}}^*(p_{l_2}) \theta(\tau_R-\tau_L)+  \bigg(\tau_R \leftrightarrow \tau_L \bigg) \bigg]\,.
\end{align}
Substituting $S_V=(\sum_{i\in v}s_i)+s_{p_{l_1},\tau_L}+s_{p_{l_2},\tau_L}$ and $s_{p,\tau_{\text{adj}}}=s_{p_{l_1},\tau_R}+s_{p_{l_2},\tau_R}$, we see by comparing with Eqn.~\eqref{Eqn:t_loop_site} that the above equation is,
\begin{align}
   D_{1_n}:=&(H)^{-4+R_V+2v+S_V+2+2} (-H)^{s_{p,\tau_{\text{adj}}}}(i)^v (-1)^{S_V+v} \bigg(\prod_{i \in v} \frac{1}{2\omega_i^2}\bigg)\tau_{R}^{-4+2+2+s_{p,\tau_\text{adj}}+R_V+S_V+v} \nonumber \\
    & \times \text{flat space expression Eqn.~\eqref{Eqn:t_loop_site}} \,. \label{Eqn:tau_loop_site}
\end{align}
and once again the flat space integral is recovered due to Eqn.~\eqref{Eqn:flat_dS_lead_prop}. This is a crucial step to recover the flat space amplitude from the correlator, since it leads to these factors being pulled out of the loop integrals. 

Similarly, the remaining time integrals in the correlator are integrated until the final (contact) time integral remains. Inspecting Eqns.~\eqref{Eqn:A_n_t_end_ans}, \eqref{Eqn:t_end_reltn}, \eqref{Eqn:t_loop_site} and \eqref{Eqn:tau_loop_site}, we see that the integrand of the final time integral will involve,
\begin{enumerate}
    \item \textbf{Factors of $H$}: Every vertex contributes $H^{-4}$, every exchange contributes $H^2$, thus we have $H^{-4V+N+2n+2I}$, where the total number of vertices is $V$, $N=\sum_{V_i}(R_{V_i}+S_{V_i})$ is the total number of spatial (and inverse laplacians) and time derivatives acting on the fields at all interaction vertices, and $I$ is the number of internal lines.
    \item \textbf{Factors of $\tau_{\text{fin}}$} : where $\tau_{\text{fin}}$ is the final time. Once again, every vertex contributes $\tau^{-4}$ and every exchange contributes $\tau^2$, so we have $\tau_{\text{fin}}^{-4V+2I+N+n}$.
\end{enumerate}
Integrating over the final power of $\tau_{\text{fin}}$ with $e^{-\omega_T \tau_{\text{fin}}}$ and keeping the leading term in the limit $\omega_T \rightarrow0$, we find,
\begin{align}
    \lim_{\omega_T\rightarrow0}D_{1_n}=(H^{\alpha+n})(-1)^{n+\sum S_V} i^n\bigg(\prod_{i \in \text{ext.}} \frac{1}{2\omega_i^2} \bigg)\frac{(\alpha!)(-1)^\alpha}{i^{\alpha+1}} \frac{\mathcal{A}_n}{\omega_T^{\alpha+1}} \,, \label{Eqn:flat_limit_D_1}
\end{align}
where $\alpha=-4V+2I+N+n$. This is the flat space limit of the $D_1$ part of the correlator, and it is straightforward to obtain the flat space limit of $D_2$ from above. Notice that $D_1$ or $\mathcal{A}_n$ can never have a total energy branch cut \cite{Bhowmick:2025mxh}, since these are produced from time integrals involving $\log (\tau)$. Regardless, due to Eqns.~\eqref{Eqn:flat_limit_D_1} and ~\eqref{Eqn:D_2}, the leading terms of the full (unrenormalized) correlator in the limit of $\omega_T\rightarrow 0$ are obtained from the flat space scattering amplitude as,  
\begin{align}
    \lim_{\omega_T\rightarrow0}D_{n}=(H^{\alpha+n})(-1)^{n+\sum S_V} i^n \frac{(\alpha!)(-1)^\alpha}{(i \omega_T)^{\alpha+1}} \bigg(\prod_{i \in \text{ext.}} \frac{1}{2\omega_i^2} \bigg) \bigg( \mathcal{A}_n +~(\sum_{j}n_j)\log \big(\frac{H}{\omega_T}\big))  \mathcal{A}_n^{(\frac{1}{\delta})}\bigg) \,, \label{Eqn:flat_space_limit}
\end{align}
where $\mathcal{A}_n^{(\frac{1}{\delta})}$ denotes the coefficient of the divergent term in the $\mathcal{O}(1/\delta)$ piece of the flat space scattering amplitude. $n_j=\frac{1}{2},-1$ for internal modes and measure respectively. Eqn.~\eqref{Eqn:flat_space_limit} is the flat space limit of the dS correlator at $1$-loop computed in dimensional regularization. Note that as far as finite non-log terms (NLf) are concerned, this limit only retrieves the leading singular terms in $D_1$. As pointed out in Appendix.~\ref{App_D_2}, the contribution from $D_2$ also produces NLf, which does not have any flat space counterpart, since this comes from the $\mathcal{O}(\delta)$ correction to modes and measure, which has no analogue in flat space. Also note that in the discussion above, we have suppressed the `$+$' notation on this correlator : which refers to the fact that $D_n^+$ ($D_n$ in Eqn.~\eqref{Eqn:flat_space_limit}) is a completely time-ordered correlator and has all vertices below $\tau=0$. There is one more diagram ($D_n^-$) at $n$-point having the same interactions and with the same order of pole at $\omega_T$, which is completely anti-time ordered, with all vertices lying above $\tau=0$. $D_n^-$ does not require a separate computation, since it is obtained from $D_n^+$ by a complex conjugation. Thus the flat space limit of the full correlator ($D_n^++D_n^-$) is obtained by taking the Real part of the right hand side in Eqn.~\eqref{Eqn:flat_space_limit}.

It is also worthwhile noting that naively Eqn.~\eqref{Eqn:flat_dS_lead_prop} suggests that an analogous flat space limit should also exist for higher loop/loop sites, since the exchange propagators for the dS correlators and scattering amplitudes are related by factors independent of loop momenta. However one should be careful since the discussion above relies heavily on solving the loop integrals by expanding the $d$-dimensional modes about $\delta=0$, and commuting this summation operator with the integral. We have demonstrated that this method computes logarithms reliably by performing a cross check with cutoff regularization. It is not immediately obvious that this method works at higher loop orders, when it would become necessary to consider $\mathcal{O}(\delta^2)$ and higher corrections from modes. Hence it is not clear whether the arguments presented above generalise straightforwardly at higher loops/loop sites. We keep this investigation for a future work.

Also, the flat space limit above is obtained from dimensional regularization. Ofcourse physical answers should not be regularization scheme dependent, however as mentioned previously mentioned, the total energy branch cut is unique to dS : there is no analogue of this branch cut in offshell amplitudes. This branch cut is a consequence of the time dependence in the background, however it has different origins in the two schemes: in cutoff regularization, it arises due to time dependence of the cutoff while in dimensional regularisation it arises from $\mathcal{O}(\delta)$ correction of modes and measure. Hence the flat space limit of this branch cut is recovered in slightly different ways for the two schemes. Hence we discuss the flat space limit for correlators being computed in cutoff regularization below. In particular, the power law divergences in $\Lambda$ arising in cutoff regularized correlators are expected to be absorbed by counter terms, while the $\log\Lambda$ is expected to turn into $\log \mu$ after renormalization. The latter tells us that the coefficient of $\log (H/\Lambda)$ term in cutoff should match the coefficient of the $1/\delta$ divergent term in dim reg, which agrees with our calculations of Sec.~\ref{Sec:Explicit_Comp_Bispectrum}. Also, since the contribution of counter terms renormalizing the 1-loop diagrams are tree level correlators computed with $3$-d modes, thus they never produce a $\log \omega_T$, i.e. all total energy branch cuts are produced in the unrenormalized correlator. Hence instead of the term $\sum_jn_j$ in Eqn.~\eqref{Eqn:flat_space_limit}, the flat space limit in cutoff will take care of all branch cuts, with $\log(H/\omega_T)$ now multiplying the coefficient of term logarithmically divergent in $\Lambda$ : $\mathcal{A}_n^{(\log \Lambda)}$. Thus for (unrenormalized) dS correlators and flat space scattering amplitudes computed in cutoff regularization, we have,
\begin{align}
    \lim_{\omega_T\rightarrow0}D_{n}=(H^{\alpha+n})(-1)^{n+\sum S_V} i^n \frac{(\alpha!)(-1)^\alpha}{(i \omega_T)^{\alpha+1}} \bigg(\prod_{i \in \text{ext.}} \frac{1}{2\omega_i^2} \bigg) \bigg( \mathcal{A}_n -\log \big(\frac{H}{\omega_T}\big)  \mathcal{A}_n^{(\log \Lambda)}\bigg) \,. \label{Eqn:flat_space_limit_cutoff}
\end{align}
Note that the flat space limit in cutoff regularization does not recover the finite non-log parts since a finite piece was dropped from the dS calculation in cutoff (See discussion below Eqn.~\eqref{Eqn:cutoff_loop_integrals}). 

In the following section, we check these relations explicitly with a few example correlators that have been previously computed in Sections.~\ref{Sec:Explicit_Comp_Bispectrum} and \ref{Sec:Complicated_Interactions}.

\subsection{Examples of flat space limit of dS correlators}
Below we consider the leading singular terms of the dS correlator in the $\omega_T\rightarrow0$ limit, and compute the scattering amplitude corresponding to the same diagram to verify the flat space limit at $1$-loop. We will keep explicit all numerical factors, as well as factors of $H$ and external energies.

\paragraph{Diagram 21} The interaction insertions in this contribution are given by $i g_4 ~\dot{\varphi}^2 \partial^{-2}\partial_i\left(\dot{\varphi}\partial_i\varphi \right)$ and $i g_3~\dot{\varphi}^2 \varphi$, hence the total number of spatial and temporal derivatives are given by $N=5$. With $V=2$, $I=2$ and $n=3$, we have $\alpha=4$. Thus the flat space limit in Eqn.~\eqref{Eqn:flat_space_limit} gives,
\begin{align}
    \lim_{\omega_T\rightarrow0}D_{n}=- H^7  \frac{4!}{\omega_T^5} \bigg(\prod_{i \in \text{1,2,3}} \frac{1}{2\omega_i^2} \bigg) \bigg( \mathcal{A}_n +~(\sum_{j}n_j)\log \big(\frac{H}{\omega_T}\big))  \mathcal{A}_n^{(\frac{1}{\delta})}\bigg) \,. \label{Eqn:flat_space_Diagram_21}
\end{align}
To verify this, we compute the \textit{off-shell} scattering amplitude corresponding to the same contraction in flat space using dimensional regularization, to get (stripping the momentum conserving delta function),
\begin{align}
    &\mathcal{A}_n =g_3 g_4 \mu^{-3\delta/2} \frac{\omega _1 }{480} \bigg[\frac{1}{\delta}\left(5 \left(2 s^2 \omega _3^2+2 s^2 \omega _{12}^2-3 \omega _3^4-3 \omega _{12}^4\right)-6 s^4\right)  \nonumber \\
    &+\bigg(10 s^2 \omega _3^2-3 s^4-15 \omega _3^4\bigg) \log \left(s+\omega _3\right)+\bigg(10 s^2 \omega _{12}^2-3 s^4-15 \omega _{12}^4\bigg) \log \left(s+\omega _{12}\right) \bigg] \nonumber \\
    & +\text{finite} \,, \label{Eqn:Diagram_21_flat}
\end{align}
while the terms of the unrenormalized dS correlator leading in the limit $\omega_T\rightarrow0$, including $\mathcal{O}(\delta)$ corrections coming from modes and measure, is given by,
\begin{align}
   &D_{n} \xrightarrow{\omega_T\rightarrow 0} g_3 g_4 \mu^{-3\delta/2} \frac{H^7}{160 \omega _1 \omega _2^2 \omega _3^2 \omega_T^5}\bigg[\bigg\{ \frac{1}{\delta}+\frac{3}{2}\log\frac{H}{\omega_T} \bigg\}\bigg(5 \big(-2 s^2 \omega _3^2-2 s^2 \omega _{12}^2+3 \omega _3^4\nonumber \\
   &+3 \omega _{12}^4\big)+6 s^4\bigg) + \bigg(-10 s^2 \omega _3^2+3 s^4+15 \omega _3^4\bigg) \log \left(s+\omega _3\right)+\bigg(-10 s^2 \omega _{12}^2+3 s^4 \nonumber \\
   & +15 \omega _{12}^4\bigg) \log \left(s+\omega _{12}\right) \bigg] +\text{NLf} \,, \label{Eqn:Diagram_21_leading}
\end{align}
where NLf denotes non log finite functions from $D_1$ and $D_2$, and the finite contribution from $D_1$, given by,
\begin{align}
    &\text{NLf}\xrightarrow{D_1}-g_3 g_4 \mu^{-3\delta/2}  \frac{H^7}{4800 \omega _1 \omega _2^2 \omega _3^2 \omega_T^5}\big(276 s^4+50 \big(-8 s^2 \omega _3^2-3 s^3 \omega _3+\omega _{12} \big(-8 s^2 \omega _{12}-3 s^3 \nonumber \\
    & +\omega _{12}^2 \left(3 s \left(3+\frac{4 i \pi  s}{\omega _{12}-\omega _3}\right)+(9-9 i \pi ) \omega _1+(9-9 i \pi ) \omega _2\right)\big)+9 s \omega _3^3+(9-9 i \pi ) \omega _3^4\big)\big) \,,
\end{align}
is related to the finite piece in the scattering amplitude of Eqn.~\eqref{Eqn:Diagram_21_flat} via the flat space limit.
Note that the unrenormalized dS correlator features unphysical logarithms, which will only be corrected once contribution from counter terms are taken into account, after which the branch cuts are arranged to give $\log(\frac{s+\omega_3}{\omega_T})$, $\log(\frac{s+\omega_{12}}{\omega_T})$ and $\log(\frac{H}{\mu})$, as in Eqn.~\eqref{Eqn:Diagram_21} (which includes the subleading pieces).

Clearly, comparing Eqns.~\eqref{Eqn:Diagram_21_flat} and \eqref{Eqn:Diagram_21_leading}, we see that the flat space limit in Eqn.~\eqref{Eqn:flat_space_Diagram_21} is satisfied.

\paragraph{Diagram 31}
This diagram could only be computed in cutoff regularization since the angular dependence coming from the vertex factors becomes intractable in $d$-dimensions. The values of $N,V,I,n,\alpha$ remain the same as in the previous case (Diagram 21). Thus the flat space limit in Eqn.~\eqref{Eqn:flat_space_limit_cutoff} gives,
\begin{align}
    \lim_{\omega_T\rightarrow0}D_{n}=-H^7 \frac{4!}{\omega_T^5}\bigg(\prod_{i \in1,2,3} \frac{1}{2\omega_i^2} \bigg) \bigg( \mathcal{A}_n -\log \big(\frac{H}{\omega_T}\big)  \mathcal{A}_n^{(\log \Lambda)}\bigg) \,. \label{Eqn:_Diagram_31_flat_space_limit_cutoff}
\end{align}

To verify this relation, we compute the flat space scattering amplitude corresponding to this contraction in cutoff.  We use the same hard cutoff $\Lambda$ that is used to regulate the loop integrals in correlators. Since the cutoff is now time independent, it is possible to perform the time integrals before the momentum integrals. This is done to avoid taking complicated fourier transforms arising from loop integrals. The calculation is then straightforward and gives,
\begin{align}
   & \mathcal{A}_n=g_3g_4 \frac{\omega _1 \omega _2}{96} \bigg[- \log (\Lambda ) \bigg(s^2 \omega _3-s^2 \omega _{12}-3 \omega _3^3+3 \omega _{12}^3\bigg) +\omega _3 \left(s^2-3 \omega _3^2\right) \log \left(s+\omega _3\right) \nonumber \\
    &+\omega _{12} \left(3 \omega _{12}^2-s^2\right) \log \left(s+\omega _{12}\right)\bigg] +\text{finite} \,, \label{Eqn:Diagram_31_flat_space}
\end{align}
while the unrenormalized dS correlator leading in the limit $\omega_T\rightarrow0$ computed in the cutoff regularization is given by,
\begin{align}
    &D_n \xrightarrow{\omega_T\rightarrow0}\frac{g_3g_4H^7}{32 \omega _1 \omega _2 \omega _3^2 \omega_T^5} \bigg[ \log \left(\frac{H}{\Lambda  \omega _T}\right) \bigg(-s^2 \omega _3+s^2 \omega _{12}+3 \omega _3^3-3 \omega _{12}^3 \bigg)  \nonumber \\
    & +\omega _{12} \left(s^2-3 \omega _{12}^2\right) \log \left(s+\omega _{12}\right)-\omega _3 \left(s^2-3 \omega _3^2\right) \log \left(s+\omega _3\right) \bigg] +\text{NLf} \,. \label{Eqn:Diagram_31_leading}
\end{align}
Once again, comparing Eqns.~\eqref{Eqn:Diagram_31_flat_space} and \eqref{Eqn:Diagram_31_leading}, we see that the flat space limit in Eqn.~\eqref{Eqn:_Diagram_31_flat_space_limit_cutoff} is indeed satisfied. Note that the NLf part of the correlator we compute does not contain all finite non log terms of the correlator (since we drop a few terms in the integrals which are not expected to produce logs; see comments below Eqn.~\eqref{Eqn:cutoff_loop_integrals}). However, even with the full NLf explicitly computed, we do not expect any flat space limits for these non-log finite terms to exist. This is because, as previously discussed, there is additional time dependence of the cutoff in dS, which has no flat space analogue. We leave this explicit check for a future work.

\paragraph{$3$-vertex Bispectrum at $1$-loop} We now check the flat space limit of the contribution to the bispectrum in Fig.~\ref{Fig:3_3v_pt}, obtained by three interaction insertions of $g_3\dot{\varphi} \partial_i \varphi \partial^{-2}\partial_i \varphi$. This interaction has $2$ temporal and spatial derivatives alongwith $1$ inverse laplacian operator, thus $N=6$ including all three vertices. With $V=3$, $I=3$, $n=3$, we have $\alpha=3$. We will compute this dS correlator (and the corresponding flat space scattering amplitude) in dimensional regularization, thus the flat space limit of Eqn.~\eqref{Eqn:flat_space_limit} gives,
\begin{align}
    \lim_{\omega_T\rightarrow0}D_{n}=-i H^6 \frac{3!}{\omega_T^4} \bigg(\prod_{i \in 1,2,3} \frac{1}{2\omega_i^2} \bigg) \bigg( \mathcal{A}_n +~(\sum_{j}n_j)\log \big(\frac{H}{\omega_T}\big))  \mathcal{A}_n^{(\frac{1}{\delta})}\bigg) \,, \label{Eqn:3_v_flat_space_limit}
\end{align}
The 3-vertex flat space scattering amplitude computed with these interactions is given by (stripped of momentum conserving delta function),
\begin{align}
    &\mathcal{A}_n = g_3^3~ \mu^{-3\delta/2}i\frac{s^3 \omega _2 ~\text{cos $\theta $}_k}{48 p_3 \left(p_3+\omega _3\right) \left(p_3+\omega _{12}\right)} \bigg[ \frac{1}{\delta} \bigg(2 p_3 \left(6 s^2+\omega _3 \left(\omega _{12}-\omega _3\right)\right)+p_3^2 \omega _T
    \nonumber \\
    &+ \left(6 s^2-\omega _3^2\right) \omega _T \bigg)  +\bigg( \frac{\left(3 s^2-p_3^2\right) \left(p_3+\omega _{12}\right) \omega _T}{p_3-\omega _{12}} \bigg) \log \left(p_3+s\right) -\bigg( \left(3 s^2-\omega _3^2\right)\nonumber \\
    &\left(2 p_3+\omega _T\right) \bigg) \log \left(s+\omega _3\right)  +\frac{2 p_3 \left(p_3+\omega _3\right) \left(\omega _{12}^2-3 s^2\right)}{p_3-\omega _{12}}\log \left(s+\omega _{12}\right) \bigg]+\text{finite} \,, \label{Eqn:3_v_flat}
\end{align}
where $\text{cos} \theta_k=\hat{k}_1\cdot\hat{k}_3$, while the leading singular terms of the unrenormalized dS correlator, including $\mathcal{O}(\delta)$ corrections from modes and measure, are given by,
\begin{align}
   & D_n \xrightarrow{\omega_T \rightarrow0} g_3^3~ \mu^{-3\delta/2} \frac{H^6 s^3 \text{c$\theta $}_k}{64 p_3 \omega _1^2 \omega _2 \omega _3^2 \left(p_3+\omega _3\right) \left(p_3+\omega _{12}\right) \omega _T^4} \bigg[ \bigg( \omega _T \left(p_3^2+6 s^2-\omega _3^2\right) \nonumber\\
   & +2 p_3 \left(6 s^2 +\omega _3 \left(\omega _{12}-\omega _3\right)\right) \bigg) (\frac{1}{\delta}+\frac{3}{2}\log\frac{H}{\omega_T}) + \frac{\left(p_3^2-3 s^2\right) \left(p_3+\omega _{12}\right) \omega _T }{p_3-\omega _{12}}\log \left(p_3+s\right)\nonumber \\
   & -\left(\omega _3^2-3 s^2\right) \left(2 p_3+\omega _T\right) \log \left(s+\omega _3\right)+\frac{2 p_3 \left(p_3+\omega _3\right) \left(\omega _{12}^2-3 s^2\right) \log \left(s+\omega _{12}\right)}{\omega _{12}-p_3}\bigg] \nonumber \\
   & +\text{NLf} \,, \label{Eqn:3_v_dS}
\end{align}
where NLf denotes non-log finite terms, and the piece coming from $D_1$, given by,
\begin{align}
    &\text{NLf} \xrightarrow{D_1} - g_3^3~ \mu^{-3\delta/2}\frac{H^6 s^3 \text{c$\theta $}_k}{192 p_3 \omega _1^2 \omega _2 \omega _3^2 \left(p_3+\omega _3\right) \left(p_3+\omega _{12}\right) \omega _T^4} \bigg(\omega _T \left(4 p_3^2+18 s^2-3 s \omega _3-4 \omega _3^2\right) \nonumber \\
    &+p_3 \left(36 s^2-3 s \omega _{12}+\omega _3 \left(-3 s+8 \omega _1+8 \omega _2-8 \omega _3\right)\right)\bigg) \,,
\end{align}
is related to the finite piece in the scattering amplitude of Eqn.~\eqref{Eqn:3_v_flat} via the flat space limit. Also, the unrenormalized correlator features unphysical $\log(\omega_T/\mu)$ terms, which are corrected once the correlator is renormalized, and the branch cuts re-arrange to give $\log \big(\frac{s+\omega_3}{\omega_T} \big)$, $\log \big(\frac{s+p_3}{\omega_T} \big)$ and $\log \big(\frac{s+\omega_{12}}{\omega_T} \big)$, as in Eqn.~\eqref{Eqn:3_v_dS_full}.

Finally, Eqns.~\eqref{Eqn:3_v_flat} and \eqref{Eqn:3_v_dS} can be compared to see that the flat space limit of Eqn.~\eqref{Eqn:3_v_flat_space_limit} is satisfied.

\section{Conclusion}
\label{Sec:Conclusion}
In this work, we have computed the $1$-loop correction to the bispectrum in slow-roll inflation, including the contributions of non-local interaction terms arising from the ADM formalism. Our analysis confirms that the analytic structure of the $1$-loop inflationary correlator obeys the diagrammatic rules previously derived in \cite{Bhowmick:2025mxh} for extracting the singularity structure of a diagram based solely on external energy flow into certain subgraphs. By inspecting the arguments of the logarithmic terms, we show that $1$-loop correlators (with both local and non-local interactions) exhibit poles and branch cuts along the negative real axis in the complex external-energy plane, exactly as predicted by these rules. Moreover, we find that these correlators contain total-energy branch cuts that combine with other branch cuts to form dilatation-invariant logarithms of comoving scale ratios.

Furthermore, we examined how flat-space scattering amplitudes are encoded in the total-energy singularities of loop-level de Sitter (dS) correlators, clarifying several subtle aspects of this correspondence. In particular, we observed that each branch cut of the dS correlator arising from the leading $\omega_T\rightarrow0$ singularity has a counterpart in the flat-space scattering amplitude—except for the total-energy branch cut, which has no flat-space analog, since this branch cut is a consequence of the time-dependent cosmological background.

The following open questions remain for future investigation:
\begin{itemize}
    \item In a cutoff regularization scheme, we encountered late-time logarithmic divergences multiplied by power-law divergences in the cutoff scale $\Lambda$. These divergences should cancel upon renormalization—consistent with the fact that the corresponding contractions produce no such divergences in dimensional regularization. It would be interesting to explicitly demonstrate the renormalization that removes these terms. Also, a detailed analysis of handling IR loop divergence in correlators in the context of non-local interactions is left for a future work.
    \item The analytic structure of cosmological correlators involving non-local interactions at higher loop orders (with multiple loop vertices) remains to be explored. Studying these cases would shed light on how the diagrammatic rules for singularity extraction generalize beyond $1$-loop diagrams. However, such calculations are extremely challenging, although significant efforts are underway \cite{Benincasa:2024lxe,Benincasa:2024ptf,Correia:2025yao}. It would also be worthwhile to investigate non-local interactions with massive or spinning fields, despite the additional complexity introduced by general Hankel functions in those computations (see \cite{Xianyu:2022jwk,Qin:2023bjk,Qin:2024gtr,Sleight:2019hfp,Sleight:2019mgd,Sleight:2020obc,Sleight:2021plv,Chowdhury:2024snc}). 
    \item It would be interesting to determine whether flat-space limits exist for correlators beyond the $1$-loop level. Our derivation of the flat-space limit at $1$-loop relied on expanding the dS mode functions, which allowed us to relate the leading terms of the dS propagators to flat-space propagators via factors independent of the loop momentum. However, interchanging this series expansion with the loop integrals in our dimensional regularization scheme is subtle, so a cross-check with a cutoff regularization is needed to verify that this method accurately captures the branch cuts. It also remains an open question whether our dimensional-regularization approach can reliably compute branch cuts at higher loop orders, which would pave the way for a straightforward generalization of flat-space limits to multi-loop correlators.

    \item In \cite{Goodhew:2020hob}, a flat-space limit was formulated at tree level in which a partial energy (external energy injected into a subdiagram) is taken to zero. The resulting correspondence involves more complicated factors than in the $\omega_T\rightarrow0$ case and depends on the exchange momenta. It would be worthwhile to investigate whether a similar partial-energy flat-space limit exists at loop level.
\end{itemize}

\section*{Acknowledgements}
We thank Suvrat Raju, Farman Ullah, Mang Hei Gordon Lee and Diksha Jain for insightful comments and discussions.
SB acknowledges support through the PMRF Fellowship of the Gov. of India. EA acknowledges support through DST INSPIRE Scholarship of the Gov. of India . DG acknowledges support from the Core Research Grant CRG/2023/001448 of the Anusandhan National Research Foundation (ANRF) of the Gov. of India.

\appendix 
\section{Evaluation of integrals involving $\log(-H \tau)$ } \label{App_D_2}
It was shown in \cite{Bhowmick:2024kld}, for interactions taken from the Goldstone action of the EFT of inflation, that $D_2$, which is the integral involving $\mathcal{O}(\delta)$ corrections from the modes and measure, is simply evaluated by pulling the $\log (-H \tau)$'s out of the time integrals by substituting $\tau \rightarrow \frac{1}{k_T}$, while the rest of the integral is simply $D_1$, yielding the relation in Eqn.~\ref{Eqn:D_2}. In particular, integrals of the type,
 \begin{align}
      \int_{-\infty}^0 d\tau_1 \int_{-\infty}^{\tau_1} d\tau_2 ~\tau_1^n \tau_2^m e^{i a \tau_1} e^{i b \tau_2} \log \left(-H \tau_{1(2)} \right)  \,,
      \label{eqn:2vertices}
 \end{align}
 with $a+b=\omega_T$, involve two cases. The first is when considering $\log(-H \tau_1)$ in the integral; this case is straightforward since the integral over $\tau_2$ is evaluated as in $D_1$, and evaluation of $\tau_1$ integral simply gives $(-\delta \log \omega_T)$ multiplied to $D_1$, alongwith non-log finite (NLf) polynomial functions. The second case considering $\log(-H \tau_2)$ is slightly more involved. The evaluation of $\tau_2$ integral now yields Exponential integral (Ei) functions of $\tau_1$, alongwith a $\log \tau_1$ term. The latter is evaluated quite easily following the logic of the previous case, once again producing $(-\delta \log \omega_T)$ multiplied to $D_1$ and NLf. The Ei functions, however, produce terms of the type $\log \left( \frac{\omega_T}{p_+ + x}\right)$ (where $x$ depends on external energies). However in \cite{Bhowmick:2024kld} it was found that these terms are finite for large $p_+$, thus they do not contribute to additional branch cuts since they vanish when $\delta$ is taken to $0$. This was a consequence of the simple structure of the interactions, resulting in the lowest power of $\tau$ to be $\sim 4$ (i.e. $m,n=2$). Larger powers of $\tau_i$'s in the integrand pull out more factors of $p_+$ in the denominator when the time integrals are computed. It turns out that $\sim \tau^4$ suffices to produce a finite integrand for large $p_+$, despite having an overall $\sim p_+^2$ dependence coming from modes, and another $\sim p_+^2$ coming from the phase space integrals. In particular, it can be shown that a $\sim \tau^3$ or larger power is required for the extra logs to vanish. However, in our case we see that the time integrals fail to pull out enough powers of $p_+$ in the denominator to counter the factors of $p_+$ coming from vertex contribution, modes and phase space integrals. This means that the extra logs can produce non vanishing contributions and must be handled with care. 

 Let us study a concrete example to demonstrate the above. We compute the contribution from $D_2$ to Diagram 21, whose $D_1$ contribution was computed in Sec.~\ref{Sec:Explicit_Comp_Bispectrum}. The loop integrand has the structure,
 \begin{align}
 &\sim \left(p_+-p_-\right){}^2 \left(p_-+p_+\right){}^2 \left(\left(p_-^2-s^2\right) \left(s^2-p_+^2\right)\right){}^{\delta /2} 
 \times \nonumber \\
 &\bigg[ \int_{-\infty}^0 d\tau_1 \int_{-\infty}^{\tau_1} d\tau_2  ~ f(\omega_i,\tau_1,\tau_2)~  e^{i(\omega_L-p_+)\tau_1} e^{i(\omega_R+p_+)\tau_2} \log (-H \tau_{1(2)}) \nonumber \\
 & +\bigg[ \int_{-\infty}^0 d\tau_2 \int_{-\infty}^{\tau_2} d\tau_1  ~ f(\omega_i,\tau_1,\tau_2)~  e^{i(\omega_L+p_+)\tau_1} e^{i(\omega_R-p_+)\tau_2} \log (-H \tau_{1(2)}) \bigg] \,, \label{Eqn:D_2_complicated}
 \end{align}
where $f$ is a function of external energies and $\tau_i$, and the dependence on $p_+,p_-$ in the first line arises from modes ($\sim p_1 p_2$) and phase space integral. $\omega_L$ and $\omega_R$ is the external energy entering the left and right vertices ($\tau_1$ and 
$\tau_2$ respectively). Notice that the vertex factor does not contribute additional factors of $p_+$ in Diagram 21. The terms in $f=\sum_i f^{(i)} \tau_1^{m_i} \tau_2^{n_i}$ are dependent on time as $(m=2,n=0)$, $(m=2,n=1)$, $(m=3,n=0)$, $(m=3,n=1)$. We will only focus on the first term, since the others are finite in the large $p_+$ limit, hence vanishing when $\delta \rightarrow 0$.

For the $\tau_1>\tau_2$ piece with $\log(-H \tau_1)$, as well as the $\tau_2>\tau_1$ piece with $\log (-H \tau_2)$, the integral is straightforward, resulting in $(-\delta \log \omega_T)$ multiplied to $D_1$ alongwith NLf. The $\tau_1>\tau_2$ piece with $\log(-H \tau_2)$ and the $\tau_2>\tau_1$ piece with $\log (-H \tau_1)$, however, also produce Exponential integrals (Ei) when the first time integral is carried out,
\begin{align}
    &\int_{-\infty}^{\tau_1} d\tau_2 ~e^{i(\omega_L-p_+)\tau_1} e^{i(\omega_R+p_+)\tau_2} ~\tau_1^2 \log(\tau_2)= \frac{i \tau_1 ^2 e^{-i \tau_1  \left(p_+-\omega _L\right)} }{p_++\omega _R} \text{Ei}\big[i \tau_1  \left(p_++\omega _R\right)\big]+... , \nonumber \\
    & \int_{-\infty}^{\tau_2} d\tau_1~e^{i(\omega_L+p_+)\tau_1} e^{i(\omega_R-p_+)\tau_2} ~\tau_1^2 \log(\tau_1)=-\frac{2 i e^{-i \tau_2  \left(p_+-\omega _R\right)} }{\left(\omega _L+p_+\right){}^3} \text{Ei}\big[i \tau_2  \left(p_++\omega _L\right)\big]+...\,,
\end{align}
where ... represents terms involving $\log(\tau)$ multiplied to a polynomial which is exactly equal to the result of the first time integral performed without the $\log$, thus the result of the final time integral follows the case of $\tau_1>\tau_2$ with $\log(-H \tau_1)$. 

Evaluation of the final time integral over the Ei function gives logarithms with $p_+$ in the arguments,
\begin{align}
    & \int_{-\infty}^0 d\tau_1 \frac{i \tau_1 ^2 e^{-i \tau_1  \left(p_+-\omega _L\right)} }{p_++\omega _R} \text{Ei}\big[i \tau_1  \left(p_++\omega _R\right)\big]=\frac{2 \log \left(\frac{\omega _L+\omega _R}{p_++\omega _R}\right)}{\left(\omega _L - p_+\right){}^3 \left(p_++\omega _R\right)} +..., \nonumber \\
    & \int_{-\infty}^0 d\tau_2 \frac{-2 i e^{-i \tau_2  \left(p_+-\omega _R\right)} }{\left(\omega _L+p_+\right){}^3} \text{Ei}\big[i \tau_2  \left(p_++\omega _L\right)\big]=\frac{2 \log \left(\frac{\omega _L+\omega _R}{\omega _L+p_+}\right)}{\left(\omega _L+p_+\right){}^3 \left(\omega _R-p_+\right)}\,, \label{Eqn:time_integrals}
\end{align}
where ... represents polynomial functions, which contribute to NLf when loop integrals are carried over. The logs, when multiplied to the prefactors involving $p_+$ in the first line of Eqn.~\eqref{Eqn:D_2_complicated} and integrated over $p_-$, yield a loop integrand which is not finite for $p_+\rightarrow \infty$. This integral must now be computed explicitly since this can give non-vanishing contribution to the singularity structure of the correlator. However, due to the particular pole structure of Eqn.~\eqref{Eqn:time_integrals}, these contributions cancel out between the $\tau_1>\tau_2$ and $\tau_2>\tau_1$ piece, as we will see below.

Adding the two cases up, the evaluation of $p_-$ integral gives,
\begin{align}
    &\sim (-s^2+p_+^2)^{\frac{\delta}{2}} ~g(s,\delta,p_+)~\bigg[ \log\omega_T \bigg\{\frac{2}{\left(\omega _L+p_+\right){}^3 \left(\omega _R-p_+\right)}-\frac{2}{\left(p_+-\omega _L\right){}^3 \left(p_++\omega _R\right)} \bigg\}\nonumber \\
    &+ \frac{2 \log \left(\omega _L+p_+\right)}{\left(\omega _L+p_+\right){}^3 \left(p_+-\omega _R\right)}+\frac{2 \log \left(p_++\omega _R\right)}{\left(p_+-\omega _L\right){}^3 \left(p_++\omega _R\right)} \bigg]  \,, \label{Eqn:pplus_integrand}
\end{align}
where $g$ is a function of $s,\delta$ and $p_+$. To be more precise, $g$ has terms dependent on $p_+^0, p_+^2$ and $p_+^4$. We only focus on the  $p_+^4$ dependence since this is the term that that blows up for large $p_+$. Also, we can simplify the above integral by writing,$(-s^2+p_+^2)^{\frac{\delta}{2}}=p_+^\delta (1-s^2/p_+^2)^{\frac{\delta}{2}}$. Expanding $(1-s^2/p_+^2)^{\frac{\delta}{2}}$ about $\delta=0$, we can discard all $\mathcal{O}(\delta^1)$ and higher terms since the integrand vanishes for $p_+\rightarrow0$.

Now naively, Eqn.~\eqref{Eqn:pplus_integrand} seems like bad news. Integrating over $p_+$ can produce non-vanishing and ``\textit{unphysical}" pieces of $\log(\omega_T/\mu)$ from the first line. However notice that,
\begin{align}
    & \int dp_+ p_+^\delta \frac{p_+^4}{(p_++a)^3(p_+-b)}=\frac{3a-b}{\delta}+\mathcal{O}(\delta^0) \,, \nonumber \\
    & \int dp_+ p_+^\delta \frac{p_+^4}{(p_+-a)^3(p_++b)}=\frac{-3a+b}{\delta}+\mathcal{O}(\delta^0) \,,
\end{align}
where we keep terms till $\mathcal{O}(1/\delta)$ since there is an overall $\delta$ multiplying Eqn.~\eqref{Eqn:pplus_integrand}. Hence the $\log(\omega_T)$ terms cancel from the $\tau_1>\tau_2$ and $\tau_2>\tau_1$ pieces. Similarly, $\log(p_++...)$ pieces will generate dilogs once the integral over $p_+$ is carried out (however these dilogs vanish since they multiply $\mathcal{O}(\delta^0)$ coefficients, and hence are overall of $\mathcal{O}(\delta)$). They can also generate $\frac{1}{\delta^2}$ terms, which will contribute to the overall $1/\delta$ divergence of the correlator. Interestingly, these terms also cancel since,
\begin{align}
    &\int dp_+ p_+^{4+\delta} \frac{\log(p_++a)}{(p_++a)^3(p_+-b)}=\frac{-3a+b}{\delta^2}-\frac{a}{\delta}+\mathcal{O}(\delta^0), \nonumber \\
    & \int dp_+ p_+^{4+\delta} \frac{\log(p_++b)}{(p_+-a)^3(p_++b)}=\frac{3a-b}{\delta^2} -\frac{b}{\delta} +\mathcal{O}(\delta^0)\,,
\end{align}
which ultimately contributes to NLf. Thus the contribution froms the $\tau_1>\tau_2$ and $\tau_2>\tau_1$ pieces once again add up to cancel the additional logarithm contributions, resulting in Eqn.~\eqref{Eqn:D_2} to hold in our case.

\section{Appendix : Diagrammatic rules to extract singularity structure} \label{Appendix:Diagrammatic_Rules}
\begin{figure}[hbt!]
\centering
\includegraphics[scale=0.6]{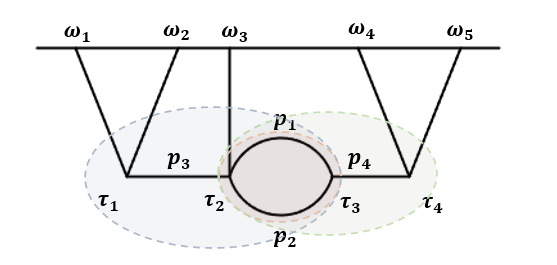}
\caption{The off-shell $5$-point function at $1$-loop, with the \textit{loop-}, \textit{left-} and \textit{right-subgraphs} shaded in red, blue and green respectively. $\tau$ denotes interaction vertices, $\omega$ denotes energy of the external legs and $p$ are internal leg momenta. }
\label{Fig:Diagrammatic_Rules}
\end{figure}

Here we briefly summarize the diagrammatic rules of \cite{Bhowmick:2025mxh}, which can be used to extract the singularity structure of any \textit{off-shell} correlation function at 1-loop with 2-sites without any computations. To proceed, one must identify all \textit{loop-}, \textit{left-} and \textit{right-subgraphs} in a diagram, as in Fig.~\ref{Fig:Diagrammatic_Rules}. These subgraphs are defined as follows.
\begin{itemize}
    \item The \textit{loop-subgraph} is the collection of the loop sites in the diagram. This is shaded in red in Fig.~\ref{Fig:Diagrammatic_Rules}.
    \item The \textit{right-subgraph} is the subgraph which includes both loop sites, such that the left loop site is an endpoint, i.e. it does not exchange with any other vertex of the subgraph except the right loop site. This is shaded in green in Fig.~\ref{Fig:Diagrammatic_Rules}.
    \item The \textit{left-subgraph} is the subgraph which includes both loop sites, such that the right loop site is an endpoint, i.e. it does not exchange with any other vertex of the subgraph except the left loop site. This is shaded in blue in Fig.~\ref{Fig:Diagrammatic_Rules}.
\end{itemize}

    The diagrammatic rules are as follows,
     \begin{itemize}
         \item If the energy flowing into the \textit{loop}-subgraph through the left and right loop site is $\mathcal{S}_L$ and $\mathcal{S}_R$ respectively, then the loop correction will feature a logarithm of $\mathcal{S}_L+s$ and $\mathcal{S}_R+s$. For e.g. in Fig.~\ref{Fig:Diagrammatic_Rules}, $\mathcal{S}_L=p_3+\omega_3$ and $\mathcal{S}_R=p_4$.
         \item If the energy flowing through the \textit{left-} (\textit{right-}) subgraph through all vertices except the right (left) loop site is $\mathcal{S}_{L_i}$ ($\mathcal{S}_{R_j}$), then the loop correction will feature a logarithm of $\mathcal{S}_{L_i}+s$ ($\mathcal{S}_{R_j}+s$). For e.g. in Fig.~\ref{Fig:Diagrammatic_Rules}, $\mathcal{S}_{L_1}=\omega_{123}$ and $\mathcal{S}_{R_1}=\omega_{45}$.
     \end{itemize}

Thus the branch cuts of the correlator in Fig.~\ref{Fig:Diagrammatic_Rules} as given by the diagrammatic rules are obtained to be $\log(p_3+\omega_3+s)$, $\log(p_4+s)$, $\log(\omega_{123}+s)$ and $\log(\omega_{45}+s)$, as well as a total energy branch cut, which is a feature of dS correlators. An explicit calculation verifies the same.

\section{Appendix : $2\rightarrow2$ scalar scattering at 1-loop in flat space} \label{App_Flat_loop}
\begin{figure}[hbt!]
\centering
\includegraphics[scale=0.4]{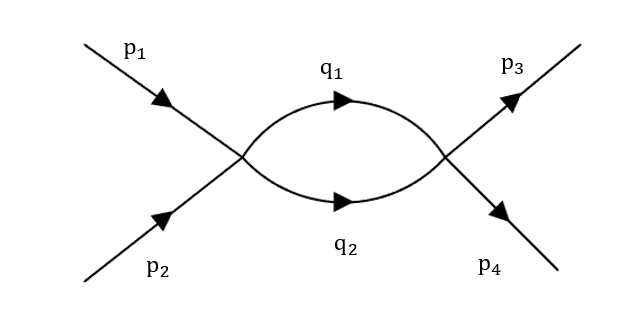}
\caption{$2\rightarrow2$ scattering of massless scalars with $\phi^4$ interaction.}
\label{Fig:1_loop_flat}
\end{figure}

Here we express the flat space scattering process in Fig.~\ref{Fig:1_loop_flat} as an integral over the loop $3$-momentum. Consider massless $\phi^4$ theory. The external particles have ($4$-)momenta $p_i=(\vec{p}_i,E_i),~ i=(1,2,3,4)$, and the loop momenta are labeled by $q_1,q_2$. Due to conservation of momenta, we have, $q_{\text{net}}=q_1+q_2=p_1+p_2=p_3+p_4$.
    
The flat space rules are framed as integral over $4$-momentum and the Feynman rules of \cite{Goodhew:2020hob} specify loops as integrals over $3$-momentum. We must integrate the former over $dq_0$ to check whether the Feynman rules used in this work reproduce the correct flat space amplitude. Following the usual flat space Feynman rules, we get the scattering process to be,
\begin{align}
    &I=\int d^4 q_1 ~\frac{1}{q_1^2+i\epsilon} \frac{1}{q_2^2+i \epsilon} \nonumber \\
    &= \int d^3\vec{q}_1 dE_{q_1} ~\frac{1}{-|\vec{q}_1|^2+E_{q_1}^2+i\epsilon} ~~\frac{1}{-|\vec{q}_2|^2+(E_1+E_2-E_{q_1})^2+i\epsilon} \,,
\end{align}

where $\vec{q}_2=\vec{p}_1+\vec{p}_2-\vec{q}_1$. The integrand has simple poles in the upper half plane at $E_{q_1}=-\sqrt{q_1^2-i\epsilon}=E_A$ , $E_{q_1}=E_1+E_2-\sqrt{|\vec{q}_2|^2-i\epsilon}=E_B$ and in the lower half plane at $E_{q_1}=\sqrt{q_1^2-i\epsilon}=E_C$ , $E_{q_1}=E_1+E_2+\sqrt{|\vec{q}_2|^2-i\epsilon}=E_D$. The integrand is finite at large $E_{q_1}$, so we can close the contour in the upper half plane. The residue at these two poles are,
\begin{align}
&\mathrm{Res}_{(E_{q_1}=E_A)} I=-\frac{1}{2|\vec{q}_1|}\frac{1}{(E_1+E_2+|\vec{q}_1|)^2-|\vec{q}_2|^2} \,, \nonumber \\
&\mathrm{Res}_{(E_{q_1}=E_B)} I=-\frac{1}{2|\vec{q}_2|}~~\frac{1}{-|\vec{q}_1|^2+(E_1+E_2-|\vec{q}_2|)^2}  \,.
\end{align}

Hence, we have,
\begin{align}
    I&=\int d^3\vec{q}_1 \bigg[ -\frac{1}{2|\vec{q}_1|}\frac{1}{(E_1+E_2+|\vec{q}_1|)^2-|\vec{q}_2|^2} -\frac{1}{2|\vec{q}_2|} \frac{1}{(E_1+E_2-|\vec{q}_2|)^2-|\vec{q}_1|^2} \bigg] \nonumber \\
    &=\int d^3\vec{q}_1 \bigg( \frac{1}{E_1+E_2+|\vec{q}_1|+|\vec{q}_2|} +\frac{1}{|\vec{q}_1|+|\vec{q}_2|-E_1-E_2|} \bigg) \frac{1}{4|\vec{q}_1||\vec{q}_2|} \,,
\end{align}
which is exactly what we get in Eqn.~\eqref{Eqn:loop_H}.

\section{Appendix : Hamiltonian for $f(\dot{\varphi})$ interactions}
\label{App_Time_dep_Hamilt}

Let us consider the Lagrangian for an interacting scalar field theory involving time derivatives,
\begin{align}
    \mathcal{L}=\frac{1}{2}(\partial \phi)^2 + \frac{g_3}{2} \dot{\varphi}^2 \varphi  \,.
\end{align}
Comparing with the ADM action Eqn.~\eqref{Eqn:cubic_action} at cubic order, we have $g_3\sim \sqrt{\epsilon}$. 

The canonical conjugate is obtained by,
\begin{align}
    \pi=\dot{\varphi}(1+g_3 \dot{\varphi} \varphi) \,.
\end{align}
To obtain the interaction Hamiltonian, we must invert the above equation to trade $\dot \varphi$ for $\pi$, and clearly this inversion will generate the interaction $\dot{\varphi}^2 \varphi$, as well as additional higher dimensional operators suppressed by the coupling. We want to investigate whether a quartic coupling will contribute to leading order. However, the quartic action in the spatially flat gauge is $\mathcal{O}(\epsilon^0)$ suppressed, and clearly at $\mathcal{O}(g_3^0)$ there are no quartic operators generated from the cubic interactions. To see this, the Hamiltonian at $\mathcal{O}(g_3^2)$,
\begin{align}
    \mathcal{H}=\frac{\pi^2}{2}+\frac{(\partial_i \varphi)^2}{2} -\frac{1}{2} g_3 \pi^2 \varphi + \frac{1}{2} g_3^2 \pi^2 \varphi^2 \,.
\end{align}

Now to extract the interaction Hamiltonian, we separate this into a quadratic $\mathcal{H}_0$ and an interaction part $\mathcal{H}_{\text{int}}$, and set $\frac{\partial {\mathcal{H}}_0}{\partial \pi}=\dot{\varphi}$. This gives, at $\mathcal{O}(g_3)$,
\begin{align}
    \mathcal{H}=\frac{\dot{\varphi}^2}{2} +\frac{(\partial_i \varphi)^2}{2} -\frac{1}{2} g_3 \dot{\varphi}^2 \varphi \,,
\end{align}
hence generating no additional quartic terms at leading order in slow roll, and we recover $\mathcal{H}_{\text{int}}=-\mathcal{L}_{\text{int}}$

\bibliography{reference.bib}

\end{document}